\documentclass[acmsmall]{acmart}

\AtBeginDocument{%
  \providecommand\BibTeX{{%
    \normalfont B\kern-0.5em{\scshape i\kern-0.25em b}\kern-0.8em\TeX}}}

\setcopyright{acmcopyright}
\copyrightyear{TODO}
\acmYear{TODO}
\acmDOI{TODO/TODO}

\acmJournal{TOMS}
\acmVolume{0}
\acmNumber{0}
\acmArticle{000}
\acmMonth{0}

\usepackage{amsmath}

\makeatletter
\def\mdseries@tt{m}
\makeatother

\usepackage[frozencache]{minted}
\newcommand{\cpp}[1]{\mintinline{cpp}{#1}}
\setminted{frame=single,fontsize=\footnotesize,breaklines,breakanywhere}

\newcommand{\R}{\mathbb{R}}
\newcommand{\X}{\mathcal{X}}
\newcommand{\I}{\mathcal{I}}
\newcommand{\N}{\mathcal{N}}
\renewcommand{\L}{\mathcal{L}}
\newcommand{\T}{\mathsf{T}}
\renewcommand{\b}{\boldsymbol}
\newcommand{\n}{\vec{n}}
\newcommand{\w}{\b{w}}
\newcommand{\dpar}[2]{\frac{\partial #1}{\partial #2}}

\begin{document}

\title{Medusa: A C++ Library for solving PDEs using Strong Form Mesh-Free methods}

\author{Jure Slak}
\authornote{Both authors contributed equally to this research.}
\email{jure.slak@ijs.si}
\orcid{0002-6405-7976}
\affiliation{%
  \institution{Jožef Stefan Institute}
  \streetaddress{Jamova cesta 39}
  \city{Ljubljana}
%  \country{Slovenia}
  \postcode{1000}
}
\affiliation{%
  \institution{Faculty of Mathematics and Physics, University of Ljubljana}
  \streetaddress{Jadranska ulica 19}
  \city{Ljubljana}
  \country{Slovenia}
  \postcode{1000}
}
\author{Gregor Kosec}
\authornotemark[1]
\email{gregor.kosec@ijs.si}
\affiliation{%
  \institution{Jožef Stefan Institute}
  \streetaddress{Jamova cesta 39}
  \city{Ljubljana}
  \country{Slovenia}
  \postcode{1000}
}

\renewcommand{\shortauthors}{Slak and Kosec}

\begin{abstract}
  Medusa, a novel library for implementation of strong form mesh-free methods,
  is described. We identify and present
  common parts and patterns among many such methods reported in the literature,
  such as node positioning, stencil selection and stencil weight computation.
  Many different algorithms exist for each part and the
  possible combinations offer a plethora of possibilities for improvements of
  solution procedures that are far from fully understood. As a consequence there
  are still many unanswered questions in mesh-free community resulting in vivid
  ongoing research in the field.
  Medusa implements the core mesh-free elements as independent blocks,
  which offers users great flexibility in experimenting with the method they
  are developing, as well as easily comparing it with other existing methods.
  The paper describes the chosen abstractions and their usage,
  illustrates aspects of the philosophy and design, offers some executions time
  benchmarks and demonstrates the application of the library on cases from
  linear elasticity and fluid flow in irregular 2D and 3D domains.
\end{abstract}




%  \includegraphics[width=\textwidth]{teaser}
%  \caption{figure caption}
%  \Description{figure description}

\maketitle

\section{Introduction}

Mesh-free (also called meshless) methods for solving partial differential equations (PDEs)
arose in 1970s and are still an active topic of research in applied mathematics today.
In mesh-free methods the computational domain is represented by a could of
points instead of a mesh of elements, as is typical for mesh-based methods.
The weak form mesh-free methods are most often analogous to the well established Finite Element
Method (FEM), while strong form methods are most often generalization of the Finite Difference
Methods (FDM).

Many strong form methods have been proposed throughout the years, starting from Smooth Particle
Hydrodynamics (SPH)~\cite{benz1990smooth}, followed by generalizations of FDM with
the Finite Point method (FPM)~\cite{onate2001finite}, the Generalized Finite Differences
method~\cite{gavete2003improvements}, and Radial basis function-generated Finite Differences
(RBF-FD)~\cite{tolstykh2003using} to name a few.
A significant
development in RBF-FD has been a recently reported by using polyharmonic RBFs
augmented with monomials~\cite{bayona2017augment} to avoid stagnation errors
and allow control over the rate of convergence. Substantial development has also been
reported in the stabilization of the method in convection dominated
regimes~\cite{shankar2018hyperviscosity}, in adaptive
solution of elliptic problems~\cite{oanh2017adaptive}, in methods for
positioning computational nodes~\cite{slak2019generation}
and in surface meshless methods~\cite{petras2018rbf,suchde2019meshfree}.

A number of mature software implementations exist for FEM, such as
deal.II~\cite{dealii}, DOLFIN (part of the FEniCS
Project)~\cite{logg2010dolfin} and FreeFem++~\cite{hecht2012new}.
Such a diverse ecosystem of general purpose implementations has not yet been developed for the
field of strong-form meshless methods. There are implementations consisting of
Matlab scripts and domain specific applications,
such as MFDMtool~\cite{milewski2013selected}, GEC\_RBFFD~\cite{bayona20153},
MFree2D~\cite{liu2002mesh}, RBFFD\_GPU~\cite{bollig2014radial}, and even a
review paper by Nguyen et al.~\cite{nguyen2008review} that specifically deals with
computer implementation, includes its own set of Matlab scripts.

Extensible, tested, documented and published general-purpose libraries for mesh-free methods
which would facilitate further research and practical applications of the field are scarce.
For older and established methods, such as SPH, high quality software packages are available,
with DualSPHyiscs~\cite{crespo2015dualsphysics} being one example.
Another such package for particle-based methods is the Aboria library~\cite{robinson2017particle}.
Two commercial mesh-free implementations are known to the authors. One is the Midas
MeshFree~\cite{midas} package, which uses the Implicit Boundary Method and a background integration
grid to perform simulations, and claims to perform ``finite element analysis''. The other is the
MESHFREE software~\cite{meshfree}, which implements the Finite Pointset
Method~\cite{tiwari2003finite} and has an impressive suite of examples. However, it focuses on
applications and not on the development of strong form mesh-free methods in general.
In 2014, Hsieh and Pan published ESFM: An essential software framework for meshfree
methods~\cite{hsieh2014esfm} which is an object-oriented C++ framework for computations using
weak-form meshless methods and claims to be the first of its kind. However, it is not publicly
available and the authors only shows examples of linear elasticity problems in the paper.
Another package to note is the RBF python package~\cite{hines2015rbf} (although not present in the
standard Python Package Index), which implements RBF interpolation and RBF-based PDE solution
techniques.

Many packages for PDE solving such as deal.II, DOLFIN,
FreeFem++, DualSPHyiscs, Aboria and ESFM libraries use the C++ programming
language. FreeFem++ implements its own extended language on top of C++ core,
while FEniCS offers Python bindings. Nonetheless, C++ seems to be the language of choice
for many such applications.
No open-source C++ library for dealing with strong form meshless methods is
known to authors. Therefore, to help further research and development in the
field of strong form meshless methods,
we present an open source C++ library Medusa (\url{http://e6.ijs.si/medusa}).

Our team started the development of Medusa library in 2015 to support our
research in the field~\cite{kosec2019weak, kosec2018local} and to ease
implementation of applied solutions~\cite{maksic2019cooling}. Over time, the interface
grew and matured, putting emphasis on modularity, extensibility and reusability. Similarly
to listed FEM libraries, it relies heavily on the C++ template system and allows the programs
to be written independently of the number of spatial dimensions with negligible run-time and
memory overhead. Special care is also taken to increase expressiveness and to be able to
explicitly translate mathematical notation into program source code. However, source code is
still standard compliant C++, which allows the user to use entirety of the C++ ecosystem.

The rest of the paper is organized as follows: a brief overview of strong-form meshless methods
is presented in section~\ref{sec:sfmm}, where the most common part of strong form meshless methods
are identified and described. This is followed by the presentation of the library in
section~\ref{sec:soft}, which also includes the relevant abstractions and
rationale behind some design decisions. Two more interesting computational examples are
presented in section~\ref{sec:applications} with measurements of execution time presented
along with comparison to FreeFem++ presented in section~\ref{sec:bench}.

\section{Strong from mesh-free methods}
\label{sec:sfmm}
Similarly to many other methods, the general parts of the solution procedure
for strong form mesh-free methods are:
\begin{enumerate}
  \item \emph{Domain discretization}: the geometry of the spatial domain is discretized,
  by placing computational nodes and finding their stencils. This part is described in
  more detail in section~\ref{sec:dd}.
  \item \emph{Differential operator discretization}: the spatial partial differential operators
  are discretized using method specific techniques. This part is described in
  more detail in section~\ref{sec:od}.
  \item \emph{PDE discretization:} The remaining time-dependent part of the PDE is discretized and
  then solved either implicitly or explicitly, with time iteration, or by only solving the implicit
  sparse system once, for elliptic problems. This part is described in
  more detail in section~\ref{sec:pded}.
\end{enumerate}

Even if the overall problem solution procedure is more complicated, and involves
coupled equations, additional physical models or non-linearities, such as in computational
fluid dynamics, the above three parts represent the core of the solution
procedure. From our experience, these parts and their components are the elements worthy of
abstraction and general implementation.

A more detailed description of the three parts is given in the following subsections,
with their respective implementations presented in sections~\ref{sec:domain-impl},
\ref{sec:approx-impl} and~\ref{sec:op-impl}.

\subsection{Domain discretization}
\label{sec:dd}
A discretization of a bounded domain $\Omega \subset \R^d$ consists of $N$ nodes
$\X = \{p_0, p_2, \ldots, p_{N-1}\}$ placed in the interior and on the boundary of
the domain. Each node is assigned a stencil (also called neighborhood or \emph{support})
consisting of some nodes near it. We will denote the size of the stencil of $i$-th
node with $n_i$ and the indices of stencil nodes with
$\I(i) = (I_{i,1}, I_{i,2}, \ldots, I_{i,n_i})$. The stencil of the $i$-th node
$\N(i)$ is the $n_i$-tuple \begin{equation}
  \N(i) = (p_{I_{i,1}}, p_{I_{i,2}}, \ldots, p_{I_{i,n_i}}).
\end{equation}
Each node should be in its own stencil, and for simplicity we assume that
it is the first one, i.e.\ $I_{i, 1} = i$ holds for all $i = 1, \ldots, N$.
Boundary nodes are assigned outer unit normals $\n_i$.

Generation of nodal distributions has often been considered
as an easy and not too relevant first step. This is partly due to the fact that
existing mesh generators could be used to generate a suitable mesh and the user
can simply discard the connectivity information~\cite{liu2002mesh}.
Besides being conceptually flawed, such approach is also computationally wasteful and does not
easily generalize to higher dimensions. Some authors even reported having difficulties to obtain
node distributions of sufficient quality~\cite{shankar2018robust}.
As a response, there are currently two known algorithm for variable density node generation
in irregular domains in arbitrary dimensions with our original algorithm~\cite{slak2019generation}
published in 2019 and another described in an arXiv preprint~\cite{van2019fast}.
Both of these algorithms are implemented in Medusa. In addition,
Medusa provides classic discretizations of basic geometric shapes, support for
gridded nodes and an ability to easily define custom node generation schemes (e.g.\ hexagonal).
Medusa also offers support for adding so-called ``ghost nodes'' to the boundary.

The remaining part of the discretization is to define the stencils, which is fully automated
and considered part of the solution procedure in nearly all meshless methods.
The most widely used type of stencils consist of some number of closest
neighbors. Besides those, balanced stencils can be used in adaptive
solutions~\cite{oanh2017adaptive}. Both approaches are implemented in Medusa,
along with the ability to only restrict the stencils to certain node types.
It is also simple to define custom stencil selection algorithms, which are not included by default,
for example visibility-based stencils~\cite{nguyen2008review}.

\subsection{Differential operator discretization}
\label{sec:od}

Most strong-form meshless approximations approximate a partial differential operator
$\L$ at a point $p$ with a linear functional $\w_{\L, p}^\T$, using an approximation of the form
\begin{equation} \label{eq:lapprox}
  (\L u)(p) \approx \sum_{j \in \I(p)} (\w_{\L, p})_j u(p_j) = \w_{\L, p}^\T \b u,
\end{equation}
where point $p$ is not necessarily one of the computational nodes.
However, the stencil indices $\I(p)$ and stencil nodes $\N(p)$ represent computational nodes.
The values $\w_{\L, p}$ are called \emph{stencil weights} or sometimes shape functions,
as a legacy terminology originating from weak-form methods.
Other approximations such as of Hermite type collocation~\cite{li2013meshless} are
also possible, but less common.

We will describe two possibilities to obtain the stencil weights $\w_{\L, p}$ which cover many
meshless formulations and are also included in Medusa by default.
The first is the generalized weighted least squares (GWLS) method,
which includes many commonly used meshless approximations, such as
SPH approximations~\cite{benz1990smooth}, Finite Point Method~\cite{onate2001finite},
Generalized Finite Difference method~\cite{gavete2003improvements}, radial basis
functions-generated finite differences (RBF-FD)~\cite{tolstykh2003using}, meshless local
strong-form method~\cite{slak2018refined}, Finite Pointset
Method~\cite{tiwari2003finite}, diffuse approximate methods~\cite{wang2012new}
and many more.

The second is a more specific radial basis functions-generated finite differences (RBF-FD)
approximation with monomial augmentation, which also offers some speed improvements.
Other custom approximation schemes can be implemented and used, such as schemes that
put additional constraints on the center weights to achieve diagonal dominance in
differentiation matrices~\cite{suchde2019meshfree}.

\subsubsection{Generalized weighted least squares}
An approximation of function $u\colon \R^d \to \R$ around $p^\ast$ is sought in the form
\begin{equation}
  \label{eq:uhat}
  \hat{u}(p) = \sum_{i=1}^m \alpha_i b_i\left(\frac{p-p^\ast}{s}\right) =
  \b{b}\left(\frac{p-p^\ast}{s}\right)^\T \b{\alpha}
\end{equation}
where $\b{b} = (b_i)_{i=1}^m$ is a set of \emph{basis functions}, $b_i\colon \R^d \to\R$,
$\b{\alpha} = (\alpha_i)_{i=1}^m$ are the unknown coefficients and $s$ is a positive scaling
factor. For simplicity we will assume that $I(p) = (1, \ldots, n)$ and $N(p) = (p_1, \ldots, p_n)$.
Note that if monomials are chosen for $b_i$, we obtain the same setup as for the standard
moving/weighted least squares (MLS/WLS) formulation~\cite{levin1998approximation}.

Using the known values $u_i$ in nearby nodes $p_i$, the error
\begin{equation}
    e_i = \hat{u}(p_i) - u_i = \b{b}\left(\frac{p_i-p^\ast}{s}\right)^\T \b{\alpha} - u_i
\end{equation}
can be computed. A weighted norm of the error vector $\b{e} = (e_i)_{i=1}^n$ is then minimized.
It can be expressed as
\begin{equation}
   \|\b{e}\|_{2,w}^2 = \sum_{i=1}^n (w_i e_i)^2 = \|W \b{e}\|_2^2 = \|W(B \b{\alpha} - \b u)\|_2^2,
\end{equation}
where $B$ is a rectangular matrix of dimensions $n \times m$ with rows containing basis function
evaluated at points $p_i$:
\begin{equation}
B =
\begin{bmatrix}
b_1\left(\frac{p_1-p^\ast}{s}\right) & \ldots & b_m\left(\frac{p_1-p^\ast}{s}\right) \\
\vdots & \ddots & \vdots \\
b_1\left(\frac{p_n-p^\ast}{s}\right) & \ldots & b_m\left(\frac{p_n-p^\ast}{s}\right)
\end{bmatrix} =
\left[b_j\left(\frac{p_i-p^\ast}{s}\right)\right]_{j=1,i=1}^{m,n} =
\left[\b{b}\left(\frac{p_i-p^\ast}{s}\right)^\T\right]_{i=1}^n,
\end{equation}
and $W$ is a diagonal matrix of weights, $W_{ii} = \omega((p_i-p^\ast)/s)$,
where $\omega\colon\R^d \to (0, \infty)$ is a \emph{weight function}.
Choosing $\omega \equiv 1$ gives the unweighted version. The arguments of $b_j$
are shifted and scaled to ensure better conditioning of matrix $B$~\cite{nguyen2008review}.

If we wanted to construct an approximant from known values of $u_i$, we could just compute
coefficients $\b \alpha$ with standard methods for solving
least square problems, such as normal equations with Cholesky decomposition, QR
decomposition or SVD decomposition.
However, to obtain an approximation of $\L|_p$, we express $\b\alpha$ in closed form
using Moore-Penrose pseudoinverse as
\begin{equation}
  \alpha = (WB)^+W\b u
\end{equation}
and substitute it in the definition~\eqref{eq:uhat} of $\hat{u}$ which
becomes \begin{equation}
  \hat{u}(p) = \b{b}\left(\frac{p-p^\ast}{s}\right)^\T  (WB)^+W\b u.
\end{equation}
The value $(\L u)(p)$ can be approximated by applying operator $\L$ to $\hat{u}$ which
gives
\begin{equation}
  (\L u)(p) \approx (\L \hat{u})(p) = (\L\b{b})\left(\frac{p-p^\ast}{s}\right)^\T (WB)^+W\b u =
  \w_{\L, p}^\T \b u,
\end{equation}
where the weights $\w_{\L, p}^\T$ are computed as
\begin{equation}
  \w_{\L, p}^\T = (\L\b{b})\left(\frac{p-p^\ast}{s}\right)^\T (WB)^+W.
\end{equation}
Note that the computation of Moore-Penrose pseudoinverse is not really necessary,
since $\w_{\L, p}^\T$ can be computed by first solving the (possibly) underdetermined system
\begin{equation} \label{eq:wbsys}
  (WB)^\T y = (\L\b{b})((p-p^\ast)/s)
\end{equation}
for $y$ and then computing $\w_{\L, p} = Wy$. System~\eqref{eq:wbsys} can be solved using
QR, SVD or any other appropriate decomposition, however, depending on $m$, $n$ and properties
of $b_j$, it can even be square and positive definite, making it possible to use
Cholesky, LDL$^\T$ or LU decompositions.

\subsubsection{Radial basis function-generated finite differences with monomial augmentation}
We again consider a partial differential operator $\L$
at a point $p$ of form
\begin{equation} \label{eq:lapprox-simple}
(\L u)(p) \approx \sum_{j = 1}^n (\w_{\L, p})_j u(p_j) = \w_{\L, p}^\T \b u,
\end{equation}
where $p_i$ are the neighboring nodes to $p$. The unknown weights in
approximation~\eqref{eq:lapprox-simple}
can be computed by enforcing equality for $n$ basis functions.
A natural choice are monomials, which
are also used in FDM, resulting in the Finite Point Method \cite{onate2001finite}.

In the RBF-FD discretization the equality is satisfied for radial basis functions $\phi_j$,
which are functions
\begin{equation}
  \left\{ \phi_j(p) =
  \phi\Bigg(\left\|\frac{p-p^\ast}{s} - \frac{p_j-p^\ast}{s} \right\|\Bigg) =
  \phi\Bigg(\frac{\left\|p-p_j\right\|}{s}\Bigg), \; j = 1, \ldots, n \right\},
\end{equation}
generated by a radial function $\phi\colon[0, \infty) \to \R$ and defined over the set of
nearby centers $p_j$. The center of the coordinate system is once again shifted to $p^\ast$ and
distances are scaled by $s > 0$ to improve conditioning.

Each $\phi_j$, for $j = 1, \ldots, n$ gives rise to one linear equation
\begin{equation}
\sum_{i=1}^{n} w_i \phi_j (p_i) = (\L \phi_j)\left(\frac{p-p^\ast}{s}\right)
\end{equation}
for unknowns $w_i$ obtained by substituting $\phi_j$ for $u$ in~\eqref{eq:lapprox}.
These equation form the following linear system:
\begin{equation} \label{eq:rbf-system}
\begin{bmatrix}
 \phi\Big(\frac{\left\|p_1-p_1\right\|}{s}\Big) &\cdots &
 \phi\Big(\frac{\left\|p_n-p_1\right\|}{s}\Big)  \\
\vdots & \ddots & \vdots    \\
 \phi\Big(\frac{\left\|p_1-p_n\right\|}{s}\Big) &\cdots &
 \phi\Big(\frac{\left\|p_n-p_n\right\|}{s}\Big)
\end{bmatrix}
\begin{bmatrix}
w_1 \\ \vdots \\ w_n
\end{bmatrix}
=
\begin{bmatrix}
(\L\phi_1)\Big(\frac{p-p^\ast}{s}\Big) \\
\vdots \\
(\L\phi_n)\Big(\frac{p-p^\ast}{s}\Big) \\
\end{bmatrix},
\end{equation}
where $\phi_j$ have been expanded for clarity.
The above system can be written more compactly as
\begin{equation} \label{eq:rbf-system-c}
A \w = \b \ell_\phi.
\end{equation}
The matrix $A$ is symmetric, and for some $\phi$ even positive
definite. Other approximation properties are also well studied~\cite{wendland2004scattered}.
Additionally, the computation up to now is the same as using GWLS with $n=m$ and $b_j = \phi_j$.

To ensure consistency up to a certain order, the computation can be augmented with monomials.
Let $q_1, \ldots, q_l$ be polynomials forming the basis of the space of $d$-dimensional
multivariate polynomials up to and including total degree $m$, with $l = \binom{m+d}{d}$.

Additional constraints are enforced by extending~\eqref{eq:rbf-system-c}
as
\begin{equation} \label{eq:rbf-system-aug}
\begin{bmatrix}
A & Q \\
Q^\T & 0
\end{bmatrix}
\begin{bmatrix}
\w \\ \b \lambda
\end{bmatrix}
=
\begin{bmatrix}
\b \ell_{\phi} \\ \b \ell_q
\end{bmatrix}, \quad
Q = \begin{bmatrix}
q_1(p_1) & \cdots & q_l(p_1) \\
\vdots & \ddots & \vdots \\
q_1(p_n) & \cdots & q_l(p_n) \\
\end{bmatrix}, \quad
\b \ell_q = \begin{bmatrix}
(\L q_1)(p^\ast) \\
\vdots \\
(\L q_l)(p^\ast) \\
\end{bmatrix}
\end{equation}
where $Q$ is a $n \times l$ matrix of polynomials evaluated at nodes $p_i$ and
$\b \ell_q$ is the vector of values assembled by applying considered operator $\L$ to
the polynomials at $p^\ast$.

%Note that the equation $Q^\T \w = \b \ell_p$ contains exactly exactness constraints for
%$q_j$. However, the introduction of parameters $\lambda_j$
%causes~\eqref{eq:lapprox} to not be exact for $\phi_i$ anymore. In fact, is was
%shown~\cite{flyer2016role} to be equivalent to the following constrained
%minimisation problem
%\begin{equation}
%\min_{\w} \left(\frac{1}{2} \w^\T A \w - \w^\T \b \ell_{\phi}\right), \text{ subject to }
%Q^\T \b
%w = \ell_q
%\end{equation}
%and parameters $\b \lambda$ can be interpreted as Lagrangian multipliers.

Weights obtained by solving~\eqref{eq:rbf-system-aug} are taken as values for $\w_{\L, p}$, while
values $\b \lambda$ are discarded.

\subsection{PDE discretization}
\label{sec:pded}
With stencil weights $\w_{\L, p}$ computed, they are mostly used in two main patterns.
The first is to explicitly approximate $(\L u)(p)$, with the field $u$ being known, such as in
explicit time iteration, during linearization of nonlinear PDEs, or simply to
obtain a derivative of the field.
The second is in implicit form, when we wish to obtain a field $u$, such that the field values
satisfy a set of linear equations.
This usually happens when solving elliptic problems or during time iteration with
at least partially implicit methods, such as Crank-Nicholson and implicit Euler's method.

Both usage patterns are described on typical examples in low-level detail in the following sections.
We judged that these patterns of spatial approximation are common enough that
suitable abstractions abstractions are offered in Medusa (see~\ref{sec:op-impl})
to avoid error-prone handling of indices, code repetition and poor readability.

\subsubsection{Explicit evaluation}
\label{sec:pded-exp}
Consider a sample time-dependent initial value problem on domain $\Omega$
\begin{align} \label{eq:expp-start}
  \hspace{4cm}  \dpar{u}{t}(p, t) &= (\L u)(p, t) && \text{ in } \Omega, \hspace{4cm} \\
  \hspace{4cm} u(p, t) &= f(p, t) && \text{ at } t=0, \hspace{4cm} \\
  \hspace{4cm}  u(p, t) &= g_d(p, t)  && \text{ on } \Gamma_d, \hspace{4cm} \\
  \hspace{4cm}  \dpar{u}{\vec n}(p, t) &= g_n(p, t)   && \text{ on } \Gamma_n, \hspace{4cm}
  \label{eq:expp-end}
\end{align}
where $\Gamma_d$ and $\Gamma_n$ are Dirichlet and Neumann boundaries, respectively, and
$f$, $g_d$ and $g_n$ are known functions. Using explicit Euler scheme in time, starting at $t=0$
with time step $\Delta t$, we define $u^k_i = u(p_i, k \Delta t)$. Time iteration using
strong form meshless approximations is performed as follows:
\begin{align} \label{eq:exp-start}
u^0_i &= f(p_i), \\
u^{k+1}_i &= u^k_i + \Delta t \left( \w_{\L, p_i}^\T u_{I(i)}^k \right), \ \text{ for internal
nodes
} p_i, \\
u^{k+1}_i &= g_d(p_i, (k+1)\Delta t), \ \text{ for Dirichlet nodes
} p_i, \\
u^{k+1}_i &= \frac{g_n(p_i, (k+1)\Delta t) - \sum_{j=2}^{n_i} u_{I_{i,j}}^k \sum_{\ell=1}^d n_\ell
(\w_{\partial_\ell, p_i})_j}{\sum_{\ell=1}^d n_\ell (\w_{\partial_\ell, p_i})_1}, \text{ for
Neumann nodes } \ p_i, \label{eq:exp-end}
\end{align}
where Neumann boundary conditions are obtained by equating the discretized version~\eqref{eq:neu}
to $g_n$ and expressing $u_i$.
Explicit discretization of Neumann boundary conditions is obtained by
approximating coordinate partial derivatives with their discrete versions
\begin{align}
 \dpar{u}{\vec n}(p_i, t) &= \sum_{\ell=1}^d n_\ell (\partial_\ell u)(p_i) \approx
   \sum_{\ell=1}^d n_\ell \w_{\partial_\ell, p_i}^\T u_{I(i)}
 = \sum_{\ell=1}^d n_\ell \sum_{j=1}^{n_i} (\w_{\partial_\ell, p_i})_j u_{I_{i,j}} \\
  &=  \sum_{\ell=1}^d n_\ell \sum_{j=1}^{n_i} (\w_{\partial_\ell, p_i})_j u_{I_{i,j}}
  =  \sum_{j=1}^{n_i} u_{I_{i,j}} \sum_{\ell=1}^d n_\ell (\w_{\partial_\ell, p_i})_j = \\
  &= u_i \sum_{\ell=1}^d n_\ell (\w_{\partial_\ell, p_i})_1 + \sum_{j=2}^{n_i} u_{I_{i,j}}
  \sum_{\ell=1}^d n_\ell (\w_{\partial_\ell, p_i})_j, \label{eq:neu}
\end{align}
where we used $I_{i, 1} = i$ and $u_{I(i)}$ is the vector of function values in stencil nodes
$u_{I(i)} = \left(u(p_j)\right)_{j \in I(i)}$.

The equations~(\ref{eq:exp-start}--\ref{eq:exp-end}) contain explicit evaluations of meshless
discretizations on known fields.
Similar expressions, containing the same explicit evaluations can be obtained for other time
discretizations or for vector functions $u$.

\subsubsection{Implicit solution}
\label{sec:implicit}
Consider a boundary value problem
\begin{align} \label{eq:impp-start}
\hspace{4cm}  \L u &= f && \text{ in } \Omega, \hspace{4cm} \\
\hspace{4cm}  u &= g_d  && \text{ on } \Gamma_d, \hspace{4cm} \\
\hspace{4cm}  \dpar{u}{\vec n} &= g_n  && \text{ on } \Gamma_n, \hspace{4cm}
\label{eq:impp-end}
\end{align}
where $\Gamma_d$ and $\Gamma_n$ are Dirichlet and Neumann boundaries, respectively, and
$f$, $g_d$ and $g_n$ are known functions. Each of the above equations is
approximated by a linear equation in corresponding computational nodes. The system of linear
equations can be written as $Mu = r$, where $i$-th row of the system corresponds to the
equation that holds in node $p_i$. Formally, the matrix $M$ and right-hand side $r$ are given by
\begin{align} \label{eq:implicit-start}
  M_{i, I_{i, j}} &= (\w_{\L, p_i})_j, \ \text{ for } j = 1, \ldots, n_i,
  & r_i &= f(p_i), && \text{ for internal nodes } p_i, \\
  M_{i, i} &= 1, & r_i &= g_d(p_i), && \text{ for Dirichlet nodes } p_i, \\
  M_{i, I_{i, j}} &= \sum_{\ell=1}^d n_\ell (\w_{\partial_\ell, p_i})_j,
  \ \text{ for } j = 1, \ldots, n_i, & r_i &= g_n(p_i), && \text{ for Neumann nodes }
  p_i.\label{eq:implicit-end}
\end{align}
Matrix $M$ is a sparse matrix with at most $\sum_{i=1}^N n_i$ nonzero entries.
Solution of the system $Mu=r$ is the numerical approximation of $u$.

The equations~(\ref{eq:implicit-start}--\ref{eq:implicit-end}) define the unknown field $u$
implicitly by using stencil weights.
Similar approximations can be obtained for vector equations, or in implicit
time stepping schemes.

\section{Software description}
\label{sec:soft}
Looking at existing finite element software packages and based on our
experience with implementing
strong-form meshless PDE solution procedures, we isolated a set of
implementation requirements:
\begin{itemize}
  \item \emph{Modularity}. Ability to change approximation, node generation, stencil selection,
  and other algorithms is of crucial importance for fast prototyping that is
  needed in research. The goal of Medusa is that different reported meshless
  methods can be rapidly constructed by using different combinations of
  provided classes.
  \item \emph{Dimension independence.} The mathematical PDE formulation is independent of the
  dimension of the problem, and we strive to conserve this property in the
  implementation as well.
  Implemented approximations, node placing algorithms and operators can be used
  in any domain dimensionality simply by changing a template parameter, e.g.
  there is virtually no difference between code for solution of problem in 2D
  or 3D, or any other dimensionality.
  \item \emph{Extensibility.} Allowing users to define their own shapes,
  approximations and operators enables wide applicability, e.g.\ implementing
  additional stabilizations such as upwind or hyperviscosity is
  straightforward.
  \item \emph{Readability.} A clear mapping from mathematical notation to code helps reduce
  errors in the code. Additionally, dealing with objects representing abstract concepts such as
  operators, vector fields and domains directly instead of matrices and lists of indices
  also helps avoid bugs.
  \item \emph{Small overhead due to the abstraction}: the run-time has
      small and often negligible overheads in comparison with ``bare-bones'' implementations.
  \item \emph{Parallelization.} When possible, parallelization can be handled internally,
  so that the program can remain relatively unchanged if the user decides for parallel execution.
  \item \emph{Ease of use.} This involves easy import and export of common file formats,
  access to examples and technical documentation.
\end{itemize}

We designed the Medusa library with above requirements in mind. The library is written
in C++ using object oriented approach and C++'s strong template system to achieve
modularity, extensibility and dimension independence. The library has no requirements,
apart from the C++ standard library and optionally the HDF5 C library~\cite{hdf5} for
reading and writing binary HDF5 files. However, we include four open-source
third-party
libraries, namely the Eigen~\cite{eigenweb} library for linear algebra,
nanoflann~\cite{nanoflann} library for spatial-search structures,
tinyformat~\cite{tinyformat} library for simple formatting and
and RapidXML~\cite{rapidxml} for XML file processing. These four libraries have been
packaged together with Medusa source code for simplicity. An external version
of Eigen can be easily used as well.

Medusa is licensed under MIT license, but the included libraries
Eigen, nanoflann, tinyformat and RapidXML are licensed under
Mozilla Public License (v.\ 2.0), BSD license, Boost Software License and
dual Boost Software license / MIT license, respectively.
The repository also includes the Google test library which
is licensed under BSD 3-Clause ``New'' or ``Revised'' License, but
is used for unit testing purposes and not necessary for core functionality.

The official website of the library is \url{http://e6.ijs.si/medusa}. The library is
developed using the \texttt{git} versioning system and the development is ongoing on
GitLab~\url{https://gitlab.com/e62Lab/medusa}. The library uses \texttt{cmake} build system
and can be used as a \texttt{cmake} submodule or as a standard standalone static C++ library.
Long compile times associated with large amounts of C++ templates are somewhat mitigated by
separating declarations from template definitions into \cpp{Medusa_fwd.hpp} and other included
files, explicitly instantiating most common class instances and linking them. If other instances are
desired, they can be explicitly instantiated or full template definitions available in
\cpp{Medusa.hpp} can be included.

Quality of implementation is ensured through continuous integration, which build the library and
runs its test suite, documentation generation tools, linters and compiles and runs
all examples. This aims to minimize the risk of regressions, stale documentation or examples and
ensures code validity, uniform code style and validity of system dependencies. The library also
includes numerous assertions, which can be disabled at compile time, that help catch errors earlier
in the debugging phase. We use Google test testing framework to develop and run over 300 tests. The
de-facto standard documentation generation tool Doxygen is used to generate the technical
documentation, which is available at \url{http://e6.ijs.si/medusa/docs}. The \texttt{cpplint} style
and code checker is used. Additionally, our wiki page is available at
\url{http://e6.ijs.si/medusa/wiki}, where more detailed explanations of examples, the
theory behind the methods, practical applications and further information about development and
potential building issues can be found.

The following section describe main modules of Medusa, dealing with domains, approximations and PDE
discretization. Almost all core classes are templated using a \cpp{vec_t} type, which contains two
essential pieces of information, the dimension of the computational domain (\cpp{vec_t::dim})
and the scalar type used for numerical computations (\cpp{vec_t::scalar_t}),
e.g.\ \cpp{float} or \cpp{complex<double>}.

\subsection{Domains}
\label{sec:domain-impl}
The main class representing domain discretizations is the \cpp{template <class vec_t>
DomainDiscretization} class, which closely resembles the description of domains discretizations
given in section~\ref{sec:dd}. It includes a list of $d$-dimensional points $p_i$,
each one has an associated \emph{type} $\tau_i$, with positive $\tau_i$ for internal nodes
and negative $\tau_i$ for boundary nodes. The boundary nodes also have their outer unit
normals $\vec n_i$ stored. Additionally, stencil indices $I(p_i)$ are stored for each point.
Stencils of varying sizes are supported.

Domain discretizations can be constructing by discretizing one of the predefined shapes,
including $d$-dimensional spheres, cubes, 2d polygons, 3d polyhedra (given by STL files),
as well as their unions, differences, translations and rotations.
Most of them support discretization of boundaries with arbitrary spacing function $h$.
For discretizations of domain interiors, two dimension independent variable density
node generation algorithms are implemented, \cpp{GeneralFill} and \cpp{GrainDropFill}, based
on~\cite{slak2019generation} and~\cite{van2019fast}, respectively.
Other node generation algorithms, such as grid-based fills and surface filling algorithms are also
available.

Two stencil selection algorithms are also available, \cpp{FindClosest}, which
constructs stencils using the indices of defined number of closest nodes, and
\cpp{FindBalancedSupport}, which also ensures that stencils are balanced around
the central node.

\begin{figure}[H]
  \includegraphics[width=0.44\linewidth]{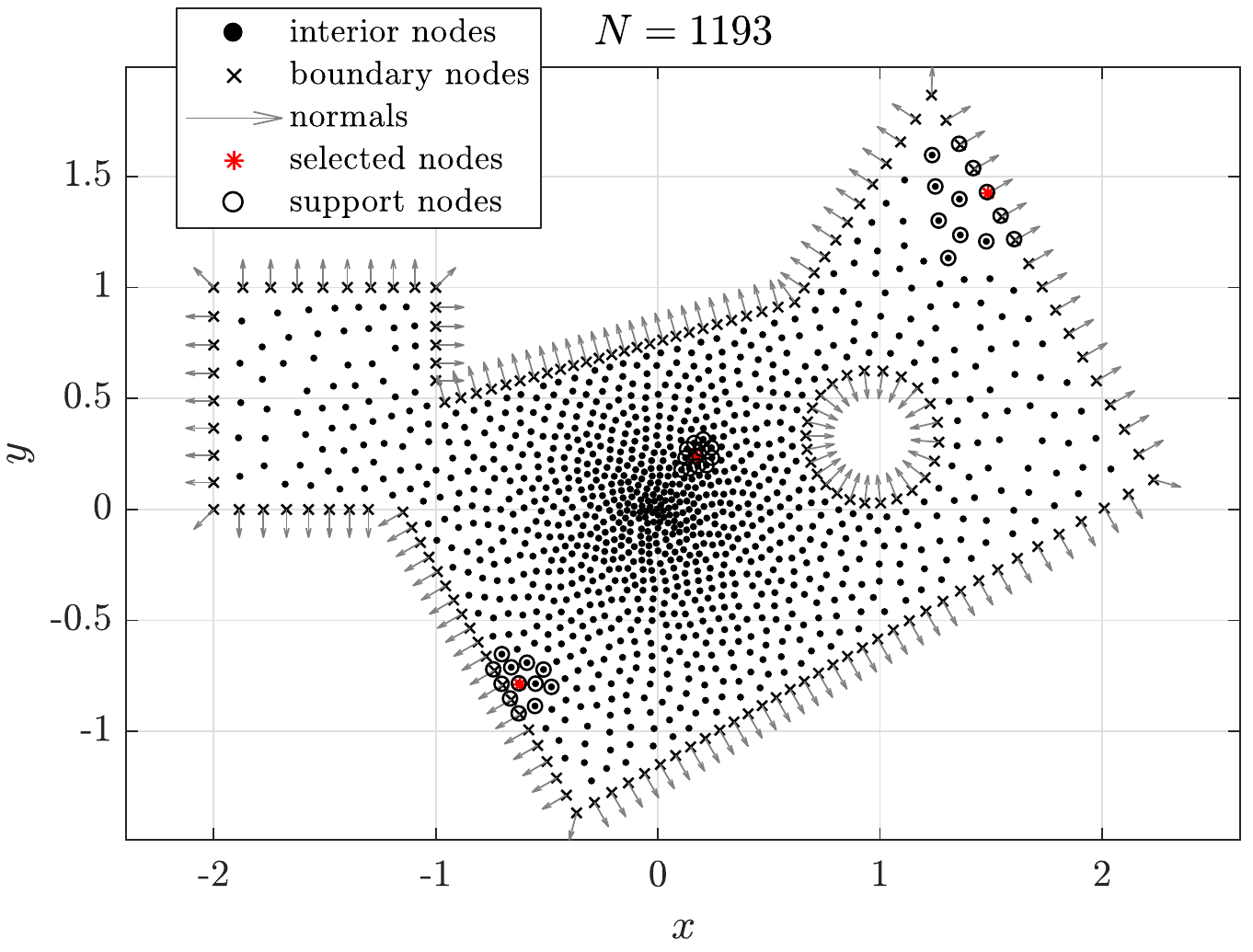}
  \includegraphics[width=0.55\linewidth]{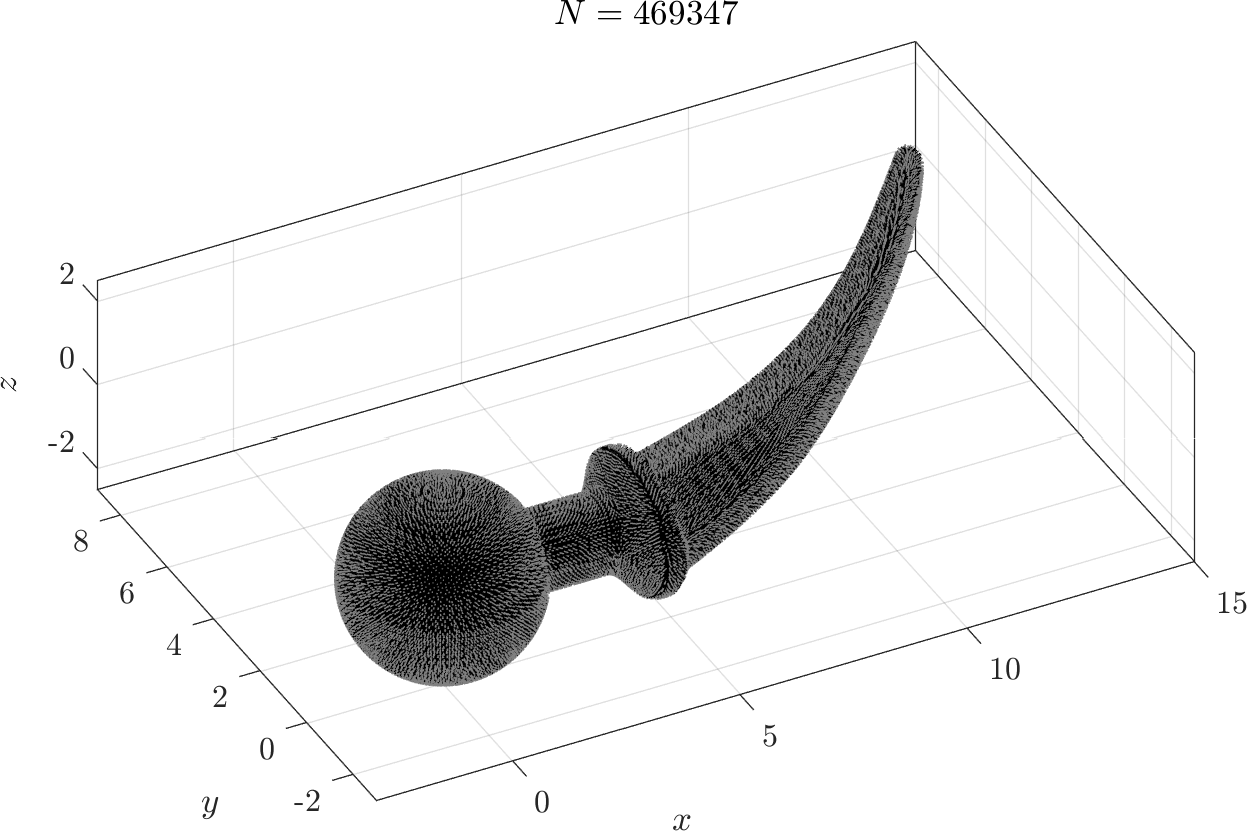}
  \caption{Domain discretizations produced by listing~\ref{lst:domains}. A few selected nodes
  are shown along with their support nodes in the left figure. The right figure shows a denser
  discretization of a STL model.}
  \label{fig:domains}
\end{figure}

Listing~\ref{lst:domains} demonstrates some of the capabilities for creating and
handling domains. Figure~\ref{fig:domains} shows the domains produced by the source code in
listing~\ref{lst:domains}. The left part shows a 2D domain with relatively coarse
variable density discretization, with interior and boundary nodes and also shows stencils
for a few selected nodes. The right part shows a uniform discretization of a 3D
model, obtained from a STL file.

\begin{listing}[H]
\inputminted[firstline=13,lastline=28,gobble=4]{cpp}{analyses/src/domain.cpp}
\caption{Construction and discretization of domains.}
\label{lst:domains}
\end{listing}

\subsection{Approximations}
\label{sec:approx-impl}
The library currently includes two approximation engines for computing
which implement the procedures described in section~\ref{sec:od}. These are
\cpp{template <class basis_t, class weight_t, class scale_t, class solver_t> class WLS},
\cpp{template <class rbf_t, class vec_t, class scale_t, class solver_t> class RBFFD},
with reasonable defaults for last few parameters.
Template parameters allow for various combinations of basis functions $b_j$, RBFs $\phi$,
weight functions $\omega$, scaling function $s$, and solvers to be used. By default,
the library includes monomial and RBF bases, Gaussian, Multiquadric, Inverse multiquadric and
Polyharmonic RBFs, three scaling functions, various weights,
and a variety of solvers included with Eigen. It is also easy for users to add their own RBFs,
weights and bases. Since templates offer a (static) version of duck typing, any class with the
interface conforming to the e.g.\ RBF concept as described in the documentation, can be used.

\begin{figure}[H]
  \centering
  \includegraphics[width=0.7\linewidth]{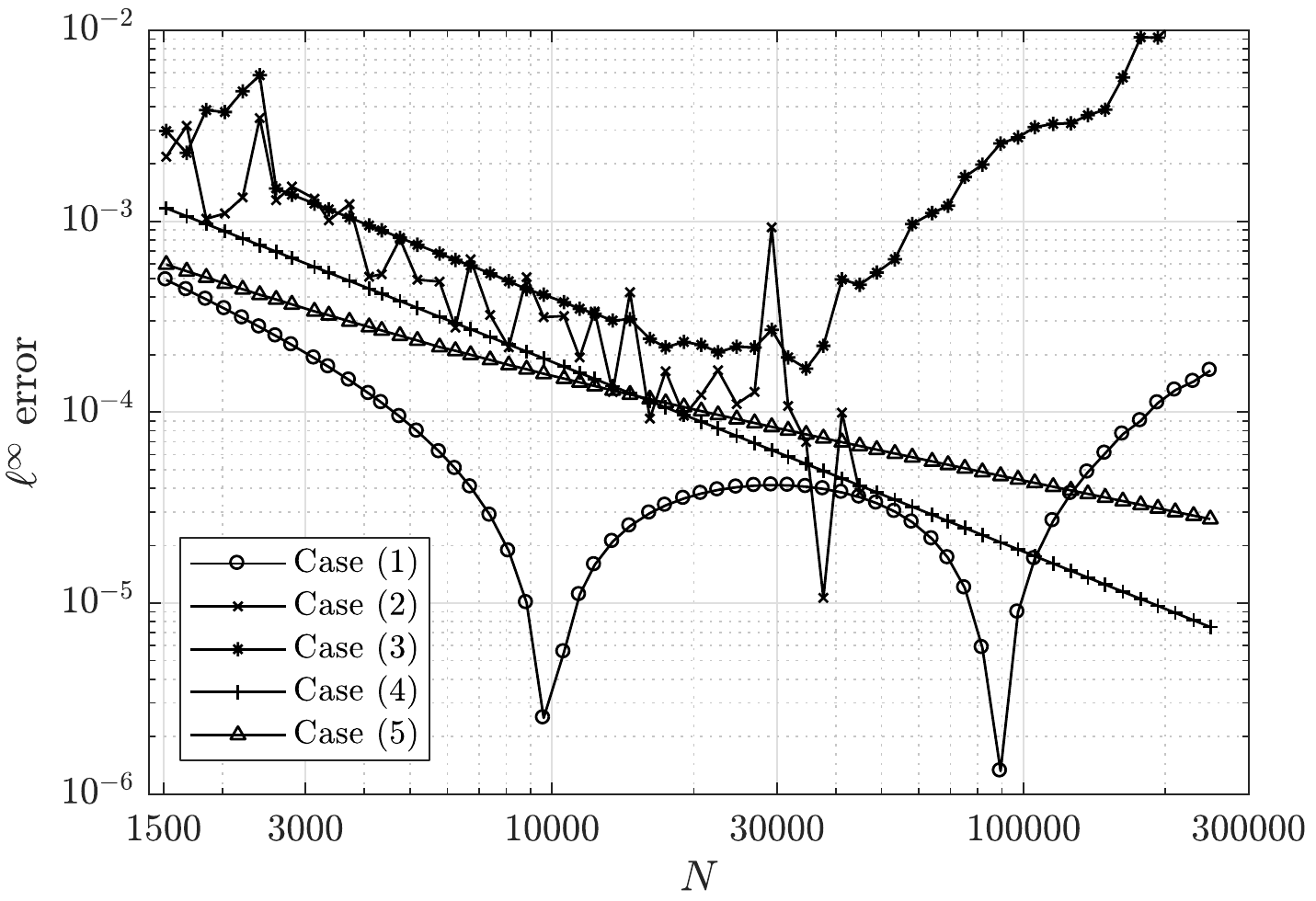}
  \caption{Error of approximating the Laplacian with different approximation setups.
  Less than 1 minute of computing time was needed to produce the data for this plot.}
  \label{fig:approx}
\end{figure}

The power of this generality is shown in Figure~\ref{fig:approx},
where errors of various approximation setups are shown.
The Laplacian operator was approximated on a regular grid $G_h$ of points with
spacing $h$ covering the unit square $[0, 1]^2$.
The error of the approximation was computed as \[
  e_h = \max_{p_i \in G_h}\left|\w_{\nabla^2\!, p_i}^\T \b u_{I(i)} - (\nabla^2 u)(p_i)\right|.
\]
The test function was chosen to be $u(x,y) = \sin(\pi x) \sin(\pi y)$.
Five different approximation setups were tested:
\begin{enumerate}
  \item RBF-FD with Gaussian RBFs $b_j(p) = \exp(\|p - p_j\|^2/\sigma^2)$, using stencil
  of $n=9$ closest nodes with no monomial augmentation, $\sigma=100$ and
  with scaling $s$ equal to the distance to the nearest neighbor.
  LU decomposition was used to solve the system for stencil weights. \label{itm:rbfgs}
  \item Like~\eqref{itm:rbfgs}, but with $\sigma=5$ and without scaling ($s=1$). \label{itm:rbfnslu}
  \item Like~\eqref{itm:rbfnslu}, but with SVD decomposition. \label{itm:rbfnssvd}
  \item GWLS with $m=5$ monomial basis functions up to order 2, $n=9$ closest nodes,
  Gaussian weight with $\sigma=1$, scaling to closest node and SVD decomposition. \label{itm:fpm}
  \item RBF-FD with polyharmonic splines $\phi(r) = r^5$ and monomial augmentation of order
  $m=2$ with $n=12$ closest nodes. \label{itm:phs}
\end{enumerate}
The definition of these setups in Medusa is shown in listing~\ref{lst:approx}.
Stencil sizes are not included, as their computation was already shown in listing~\ref{lst:domains}.

\begin{listing}[H]
  \inputminted[firstline=9,lastline=13,gobble=4]{cpp}{analyses/src/approx_sample.cpp}
  \caption{Definition of various important approximations.}
  \label{lst:approx}
\end{listing}

These setups present some of the problems and answers in meshless strong form methods in recent
years. The question of choice of the shape parameter for RBFs is a long standing one, since the
shape parameter often presents a trade-off between accuracy and the condition number of the
matrix~$A$~\cite{wendland2004scattered}.
Case~\eqref{itm:rbfnslu} exhibits the expected behavior that Gaussian approximations converge
until the condition number is too high, and numerical errors become predominant. Jagged behavior
can be smoothed by using SVD decomposition, however the overall outcome is the same.
%Stable algorithms specifically for Gaussian RBFs exist~\hl{[?]}, but are rather expensive and
%complicated.
A simple remedy for this instability is to scale the shape parameter (or the space) to keep the
condition number constant. This solves the problems with numerical instability, but causes the
approximation to diverge in a characteristic fashion with two local minimums~\cite{bayona2010rbf}.
This lack of convergence is also often called divergence due to ``stagnation errors''.
Two more convergent cases are included, one is the Finite point method (Case~\eqref{itm:fpm}),
which achieves similar
behavior and accuracy to FDM~\cite{onate2001finite} and another is RBF-FD using PHS augmented with
monomials (Case~\eqref{itm:phs})~\cite{bayona2017augment}, where accuracy and convergence order can
be easily controlled through augmentation.

\subsection{Operators}
\label{sec:op-impl}
This module defines one of the core functions of the library, which takes
a domain discretization with nodes $p_i$, an approximation engine and a list of operators
$(\L_1, \ldots, \L_\ell)$
and computes and stores stencil weights $(\w_{\L_j, p_i})_{i=1, j=1}^{N, \ell}$,
for all operators and all computational nodes in the domain.
These weights are stored in a \cpp{ShapeStorage} class.

The library supports computing shapes for first and second derivatives, as well as for the
Laplacian operator. Note that this allows for construction of arbitrary
second order operators as
\begin{align}
 \w_{\L, p} = \sum_{1 \leq |\alpha| \leq 2} a_\alpha(p) \w_{\partial_\alpha, p}, \text{ for }
 \L = \sum_{1 \leq |\alpha| \leq 2} a_\alpha(p) \dpar{}{x^\alpha},
 \label{eq:gen-op}
\end{align}
where $|\alpha| = \sum_{i=1}^d \alpha_i$ and $ \dpar{}{x^\alpha} = \dpar{^{|\alpha|}}{x^\alpha_1
\cdots x^{\alpha_d}}$ are the standard multiindex notations.
This would also cover the Laplacian operator, however, equation~\eqref{eq:gen-op} is not necessarily
the most efficient nor the most numerically stable way of computing the Laplacian for certain basis
functions. User defined operators are supported as well, with the only requirement being that
the user implements application of the operator for a class of basis functions that is used in their
code. Our examples include solving the biharmonic equation to demonstrate this extensibility.

The \cpp{ShapeStorage} class stores the computed weights for a chosen set of operators.
These shapes can be used to implicitly express or explicitly compute $\L u$, as described in
sections~\ref{sec:pded} and its subsections.

For given scalar or vector field $u$, we directly support most common scalar
and vector operators, such as coordinate derivatives of first and second order,
Laplacian, gradient, divergence, gradient of divergence, directional derivatives, as
well as any user defined operators.

Two examples of PDE solutions will be given in this section, to illustrate the
functionality of the library for explicit and implicit solving. Special effort
was put into readability of the solution procedures, to give the user
a direct mapping from the mathematical solution procedure to the source code.

\subsubsection{Explicit operators}
\begin{figure}[H]
  \centering
  \includegraphics[width=0.44\linewidth]{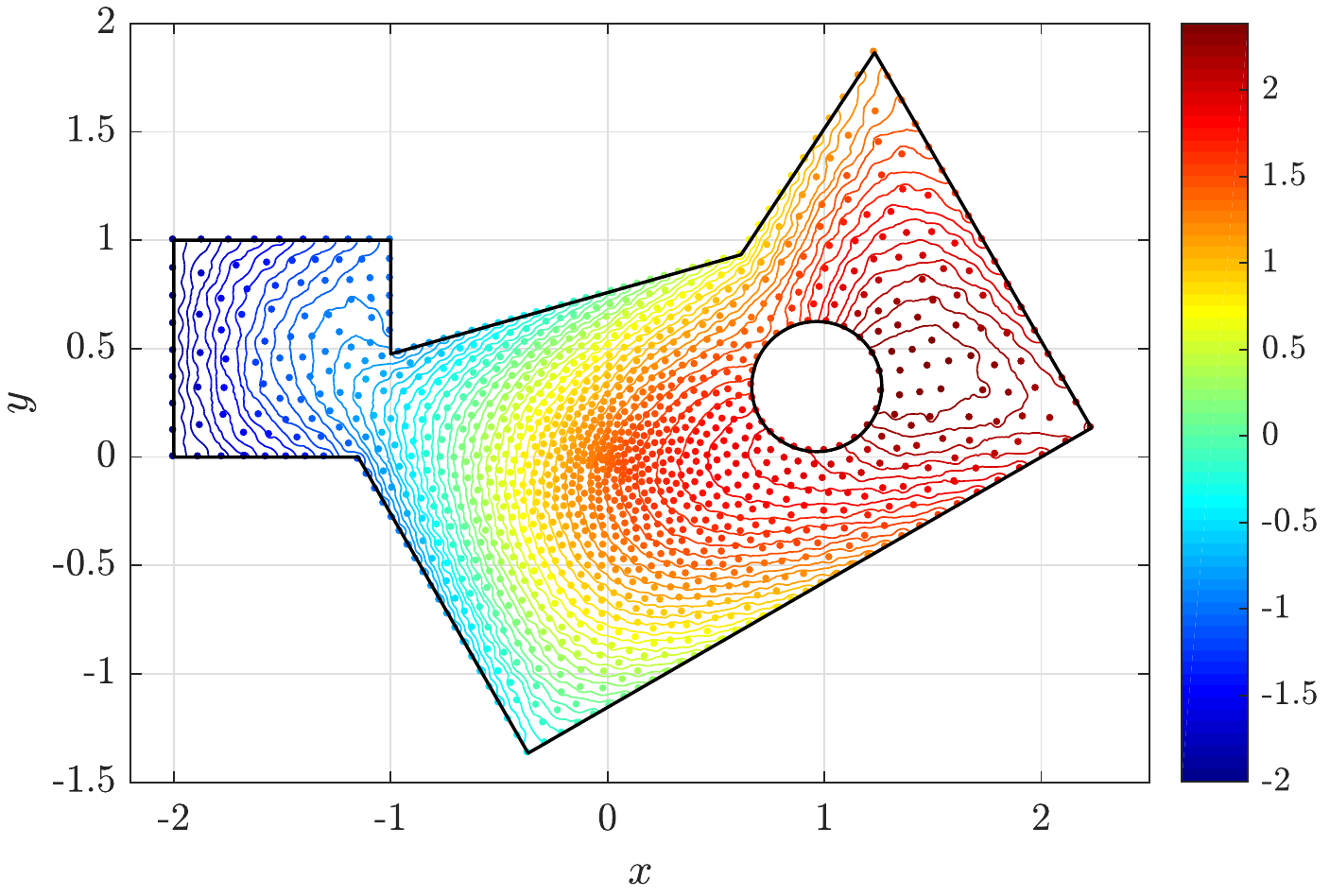}
  \includegraphics[width=0.55\linewidth]{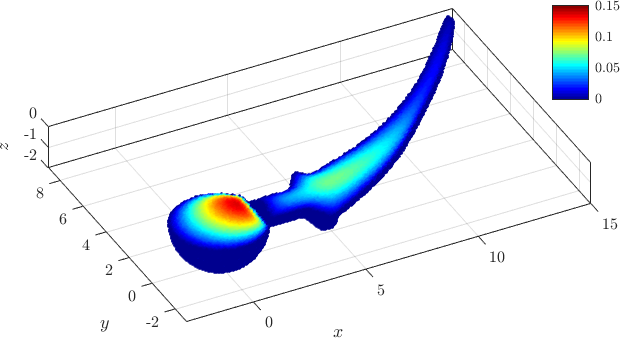}
  \caption{Solution of the heat
  equation~(\ref{eq:heat-start}--\ref{eq:heat-end}) on the left
           and convection-diffusion problem~\eqref{eq:pois} on the right.}
  \label{fig:pde-examples}
\end{figure}

Consider the problem of type~(\ref{eq:expp-start}--\ref{eq:expp-end}):
\begin{align} \label{eq:heat-start}
  \hspace{4cm} \dpar{u}{t}(x, y, t) &= \nabla^2 u + 5 && \text{ in } \Omega, \hspace{4cm} \\
  \hspace{4cm} u(x, y, t) &= 0 && \text{ at } t=0, \hspace{4cm} \\
  \hspace{4cm} u(x, y, t) &= x  && \text{ on } \Gamma_d, \hspace{4cm} \\
  \hspace{4cm} \dpar{u}{\vec n}(x, y, t) &= 0  && \text{ on } \Gamma_n, \hspace{4cm}
  \label{eq:heat-end}
\end{align}
on the 2D domain $\Omega$ constructed in listing~\ref{lst:domains},
where $\Gamma_n$ is the inner circle boundary and $\Gamma_d$ the outer boundary.
The problem is solved in listing~\ref{lst:demo2d} and the solution procedure
follows~(\ref{eq:exp-start}--\ref{eq:exp-end}). The solution is shown on the left side of
Figure~\ref{fig:pde-examples}.

Listing~\ref{lst:demo2d} begins after the domain has been constructed and the sets of indices
\cpp{interior}, \cpp{boundary} and \cpp{circle}, corresponding to the
interior, outer boundary in inner boundary nodes, respectively, have been defined.
The \cpp{computeShapes} method computes shapes for Laplacian and first derivatives,
which are then stored. The explicit operators \cpp{op} are a collection of
methods, that implement the spatial parts of the formulas~(\ref{eq:exp-start}--\ref{eq:exp-end})
from section~\ref{sec:pded-exp} and greatly help with the readability of the solution procedure.

\begin{listing}[H]
  \inputminted[firstline=27,lastline=51,gobble=4]{cpp}{analyses/src/demo2d.cpp}
  \caption{Solving the heat equation~(\ref{eq:heat-start}--\ref{eq:heat-end}) explicitly.}
  \label{lst:demo2d}
\end{listing}

\subsubsection{Implicit operators}
Consider a boundary value problem of type~(\ref{eq:impp-start}--\ref{eq:impp-end}):
\begin{equation} \label{eq:pois}
  -2 \nabla^2 u + 8\;(2, 1, -1) \cdot \nabla u = 1 \text{ in } \Omega,
   \quad u = 0 \text{ on }\partial \Omega,
\end{equation}
where $\Omega$ is the right domain in Figure~\ref{fig:domains}.
The listing~\ref{lst:demo3d} shows the source code needed to solve the problem implicitly, as
described in section~\ref{sec:implicit}.

\begin{listing}[H]
  \inputminted[firstline=12,lastline=29,gobble=4]{cpp}{analyses/src/demo3d.cpp}
  \caption{Solving convection-diffusion equation~\eqref{eq:pois} implicitly.}
  \label{lst:demo3d}
\end{listing}

After computing the weights, the appropriately allocated sparse matrix and right side are assembled.
The implicit operators \cpp{op.lap} hold a reference to the matrix and the right hand side
and them with the appropriate weights, taken from \cpp{storage}, implementing
formulas~(\ref{eq:implicit-start}--\ref{eq:implicit-end}). This is done to improve readability;
note the similarity between the line of the source code, which
defines the equation in the interior, and the equation~\eqref{eq:pois}.
The implicit system can also be amended manually, if desired.
The intuitive mathematical syntax supports expressions of form
$\sum_\ell \alpha_\ell \L_\ell u = r$, where 0th, 1st and 2nd derivatives are supported
for $\L_\ell$, as well as directional derivatives, gradients of divergence, Laplacian,
and any user defined operators.
Another benefit of this system is that (in \cpp{DEBUG} mode) checks are performed that
the operators added together always write to the same matrix row, to avoid indexing errors.

Overall, the abstractions for implicit and explicit operators are in our opinion one of the best
features of Medusa library. They allow the user to think in terms of
field and operators, instead in terms of arrays and indices, which are
much more error prone. This shift has been present in FEM implementations for a while,
with FreeFem++, Fenics and deal.II implementing these types of abstractions,
however it has not been noted in the strong form community and
finite difference codes are often riddled with poorly readable
discretization code clouding the overall problem solution procedure.
Munthe-Kaas and Haveraaen in 1996 introduced the concept of coordinate free
numerics~\cite{munthe1996coordinate}, which encompasses this idea, and Medusa
has been investigated in this direction as well~\cite{slak2018mipro}.

\subsection{Miscellaneous}
There are a few additional modules in the library that simplify its usage or offer
often needed utilities. The ``types'' module implement nicer interfaces
and additional functionality to types used to represent (physical) vectors, scalar fields,
vector fields and containers, while retaining full compatibility with Eigen.
Input and output capabilities from and to CSV, XML and HDF file formats are supported.
Some basic integrators for solving ODEs, such as RK4, are also included.

%\hl{A diagram of library organization?}
%\cm{bi bil dobrodosel, da se vidi kompleksnost paketa}

\section{Examples}
\label{sec:applications}
Plenty of examples are included in the project's repository and a tutorial
for solving the Poisson equation is available on the website.
The examples include many different setups for
solving Poisson boundary value problems, which are used to demonstrate different features.
Other examples include solving problems from electromagnetic scattering, which includes support for
complex numbers, Navier-Stokes equations for fluid simulation, problems from linear elasticity and
simulation of wave propagation. The instruction on compiling and running these examples are
available on the wiki and from the README in the examples folder.

In this paper, we include examples from linear elasticity and fluid mechanics.

\subsection{Linear elasticity}

Small displacements in an isotropic homogeneous linearly elastic material under stress are
described by Cauchy-Navier equations
\begin{equation}
(\lambda + \mu) \nabla (\nabla \cdot \vec{u}) + \mu \nabla^2 \vec{u} = \vec{f},
\end{equation}
where $\vec{u}$ are unknown displacements, $\vec{f}$ is the loading
body force, and $\lambda$ and $\mu$ are material constants, called
Lam\'{e} parameters. The stress tensor $\sigma$ is computed as
\begin{equation}
\sigma = \lambda \operatorname{tr}(\varepsilon) I + 2 \mu \varepsilon,
 \quad \varepsilon = \frac{\nabla \vec{u} + (\nabla \vec{u})^\T}{2},
\end{equation}
where $I$ is the identity tensor.

We consider a beam of dimensions $L \times W$ in 2D and $L \times W \times T$ in 3D,
occupying the area $[0, L] \times [0, W]$ in 2D and $[0, L] \times [0, W] \times [0, T]$ in 3D.
The beam is fixed on the side with the first coordinate equal to 0, experiences
a downwards traction of size $F$ on the side with the fist coordinate equal to $W$ and
zero traction elsewhere.

\begin{figure}[h!]
  \includegraphics[width=0.45\linewidth]{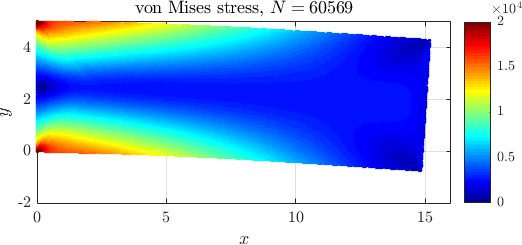}
  \includegraphics[width=0.45\linewidth]{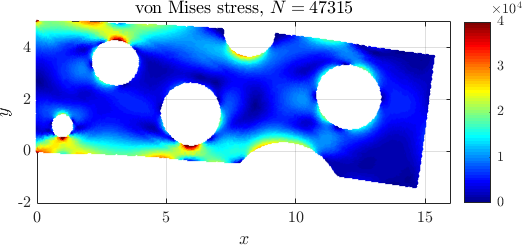}
  \includegraphics[width=0.45\linewidth]{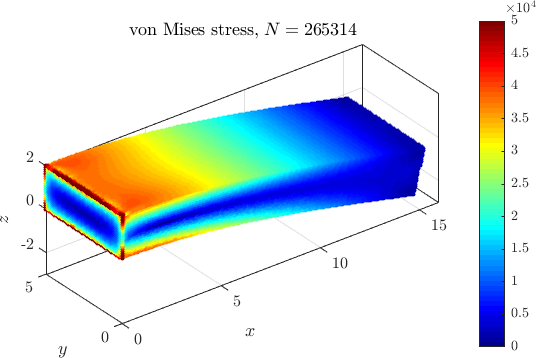}
  \includegraphics[width=0.45\linewidth]{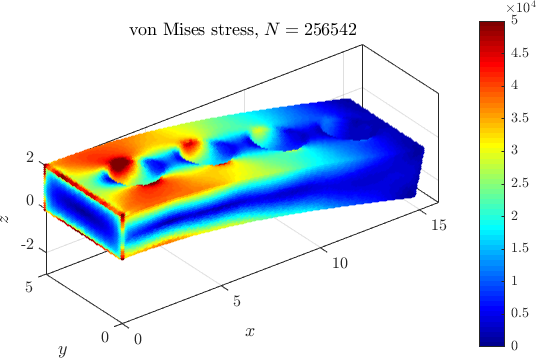}
  \caption{Cantilever beams with and without cavities in 2D and 3D.
     Displacements are multiplied by a factor $10^5$ in 2D and $5\cdot 10^4$ in 3D.}
  \label{fig:beams}
\end{figure}

Note that this is not the classical Timoshenko beam, although the library was also tested against
that problem~\cite{slak2018refined}. Additionally, some cavities (also with no traction boundary
conditions) were added to the domain. The problems were solved for $L = 15$, $W = 5$, $T = 2$,
with $E = 72.1 \cdot 10^9$, $\nu = 0.33$ and $F = 1000$. Polyharmonic radial basis function on 25
nearest nodes with monomial augmentation od 2nd order were used both in 2D and 3D.
The results are shown in Figure~\ref{fig:beams}, colored according to von Mises stress.

\subsection{Simulation of natural convection}
\begin{figure}[h!]
  \includegraphics[width=0.45\linewidth]{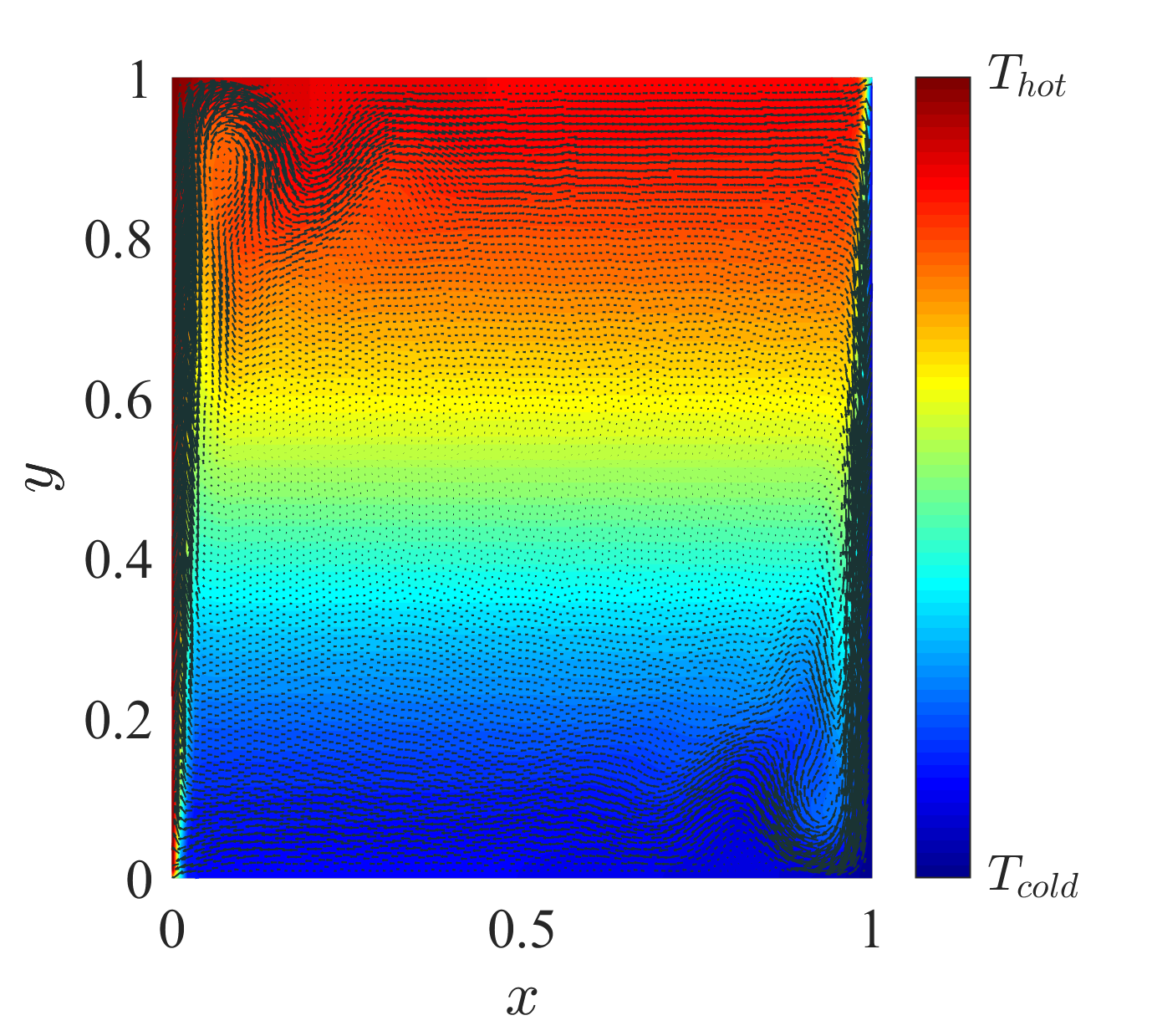}
  \includegraphics[width=0.45\linewidth]{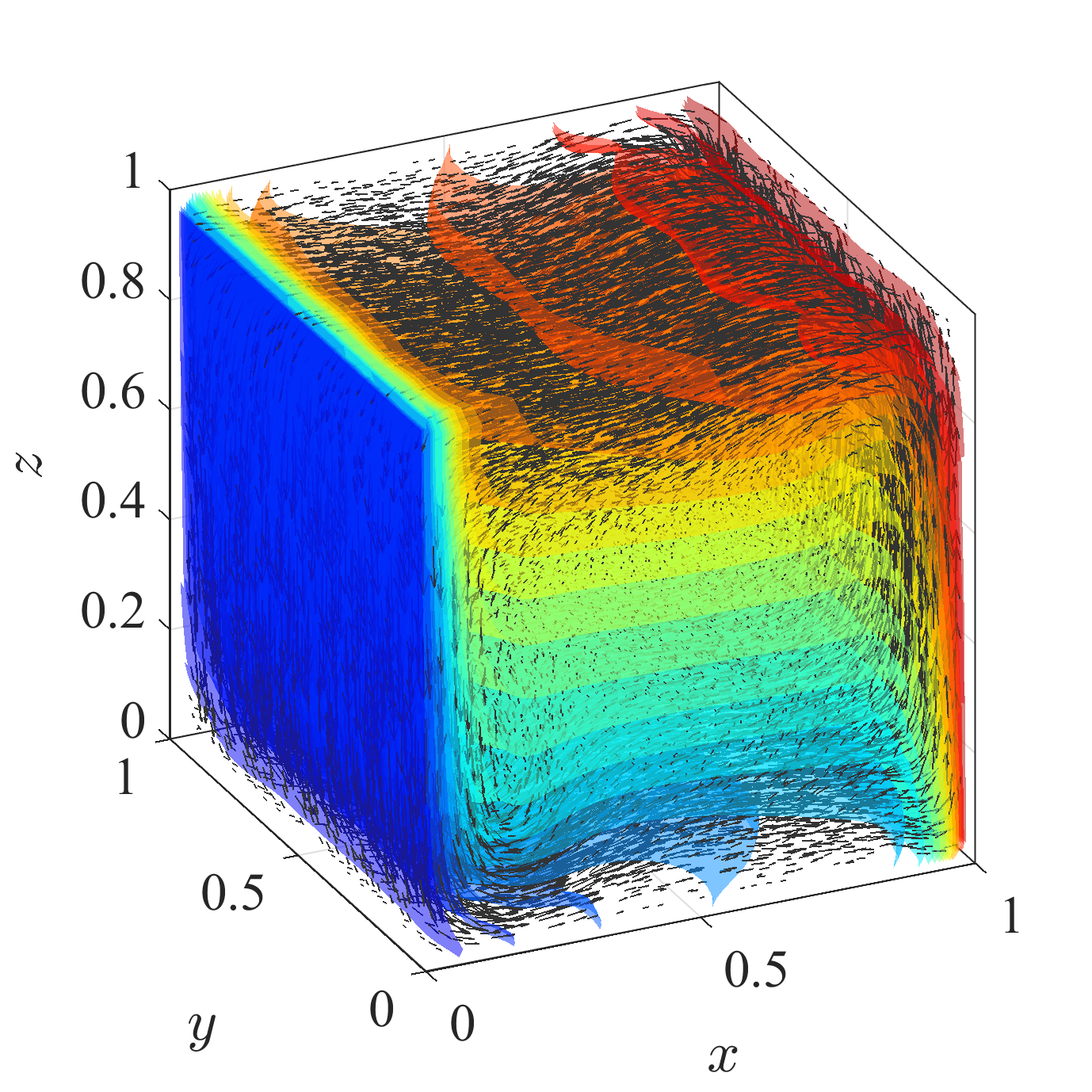}
  \includegraphics[width=0.45\linewidth]{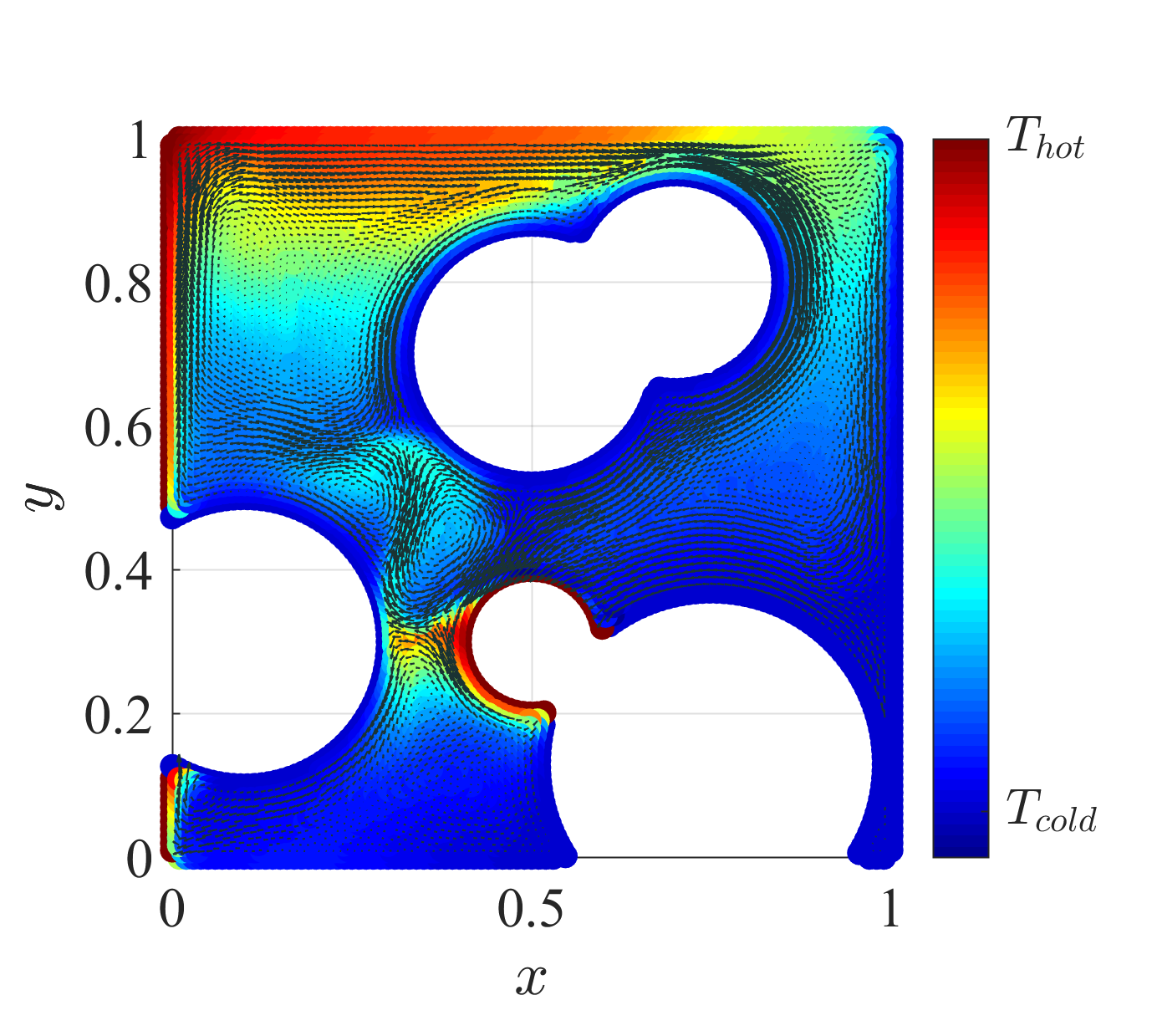}
  \includegraphics[width=0.45\linewidth]{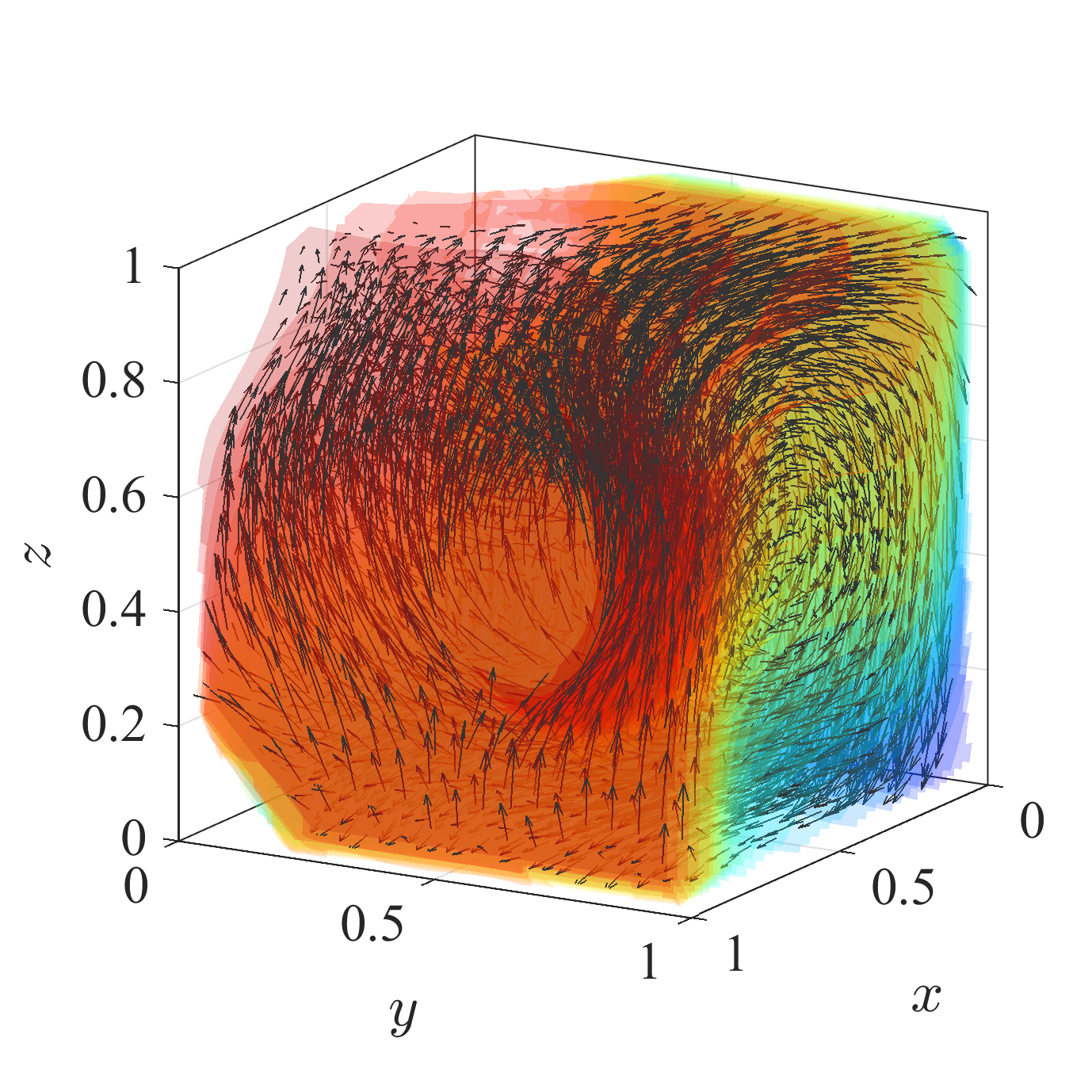}
  \caption{Solution of natural convection problem for Ra=$10^8$ in 2D (top
  left), Ra=$10^6$ in 3D (top right), and on irregular 2D and 3D domains
  (bottom row).}
  \label{fig:thermo}
\end{figure}

The natural convection problem is governed by coupled Navier-Stokes, mass
continuity and heat transfer equations
\begin{align}
\dpar{\b v}{t} + (\b v \cdot \nabla) \b v &= -\frac{1}{\rho} \nabla p +
\frac{\mu}{\rho} \nabla^2
\b v + \frac{1}{\rho} \b b, \\
\nabla \cdot \b v &= 0, \\
\b b &= \rho (1-\beta(T - T_{\text{ref}})) \b g, \\
\dpar{T}{t} + \b v \cdot \nabla T  &= \frac{\lambda}{\rho c_p} \nabla^2 T,
\end{align}
where $\b v(u,v,w)$, $p$, $T$, $\mu$, $\lambda$, $c_p$, $\rho$, $\b g$,
$\beta$, $T_{\text{ref}}$ and
$\b b$ stand for velocity, pressure, temperature, viscosity, thermal
conductivity,
specific heat, density, gravitational acceleration, coefficient of thermal
expansion,
reference temperature for Boussinesq approximation, and body force,
respectively. The problem is defined on a unit square domain with vertical
walls kept at constant different temperatures, while
horizontal walls are adiabatic. In generalization to 3D front
and back walls are also assumed to be adiabatic~\cite{wang2017numerical}. The
problem is solved with implicit time stepping and projection method for pressure-velocity
coupling~\cite{slak2019generation}. Results in terms of velocity and temperature contour plots
are presented in Figure~\ref{fig:thermo} for Prandtl number $0.71$ and Rayleigh
numbers $10^8$ in 2D and $10^6$ in 3D, respectively.
More details about the solution procedure and results
can be found in~\cite{slak2019generation}.

\section{Benchmarks}
\label{sec:bench}
While the design of Medusa is mainly focused on modularity and extensibility,
we still take care that the implementation is reasonably efficient.
To this end, we compare the performance of Medusa with the mature FreeFem++ library for
solving PDEs. Note that we will be comparing two different methods for solving
PDEs, which by themselves have different complexity,
and it is not the purpose of this measurements to compare the methods, nor the quality of
implementations. We simply wish to establish that Medusa execution times are in the same
ballpark as the FreeFem++ ones for the same problem.

The comparison is done on the Poisson boundary value problem
\begin{equation} \label{eq:time-prob}
  -\nabla u = f \text{ in } \Omega, \; u = u_0 \text{ on } \partial\Omega,
\end{equation}
for $u_0(x) = \prod_{i=1}^d \sin(\pi x_i)$ and $f = -\nabla^2 u_0$ on
$\Omega = B(0, 1) \setminus B(0, 1/2)$ in 2D and 3D.
Medusa implementation uses RBF-FD with PHS on $n=9$ and $n=35$ closest nodes in 2D and 3D,
respectively. FreeFem++ implementation solves the corresponding variational formulation using P1
elements. The problem itself and the FreeFem++ code were taken from FreeFem++'s own example suite.

FreeFem++ and its dependencies were compiled from source, as was Medusa.
Both implementations were run single-threaded on a laptop computer with
\texttt{Intel(R) Core(TM) i7-7700HQ CPU @ 2.80GHz} processor with
16 GB of DDR4 RAM. Each time measurement was repeated 9 times and the median
values are shown, with error bars showing standard deviation of the measurements.

\begin{figure}
  \includegraphics[width=0.49\linewidth]{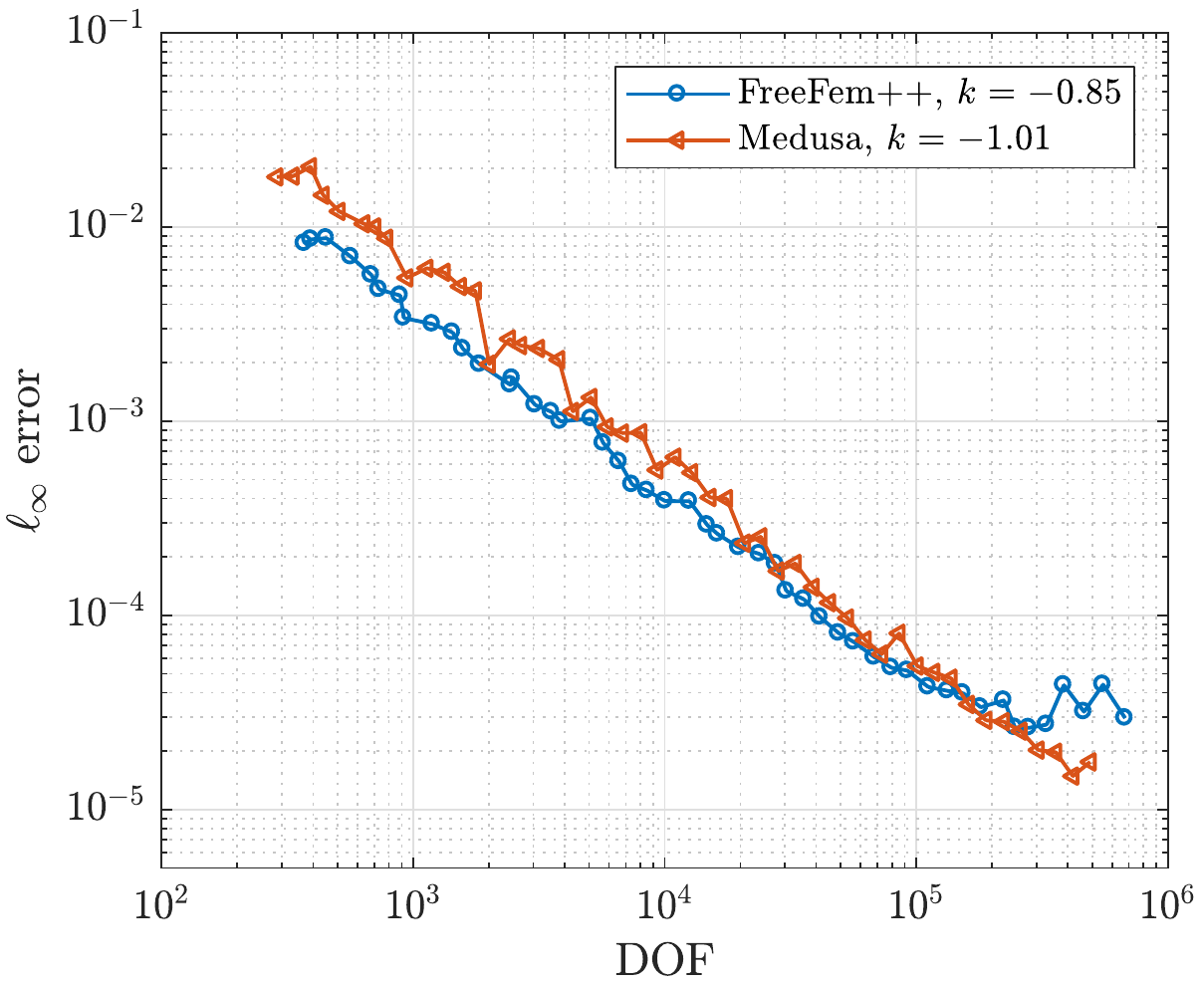}
  \includegraphics[width=0.49\linewidth]{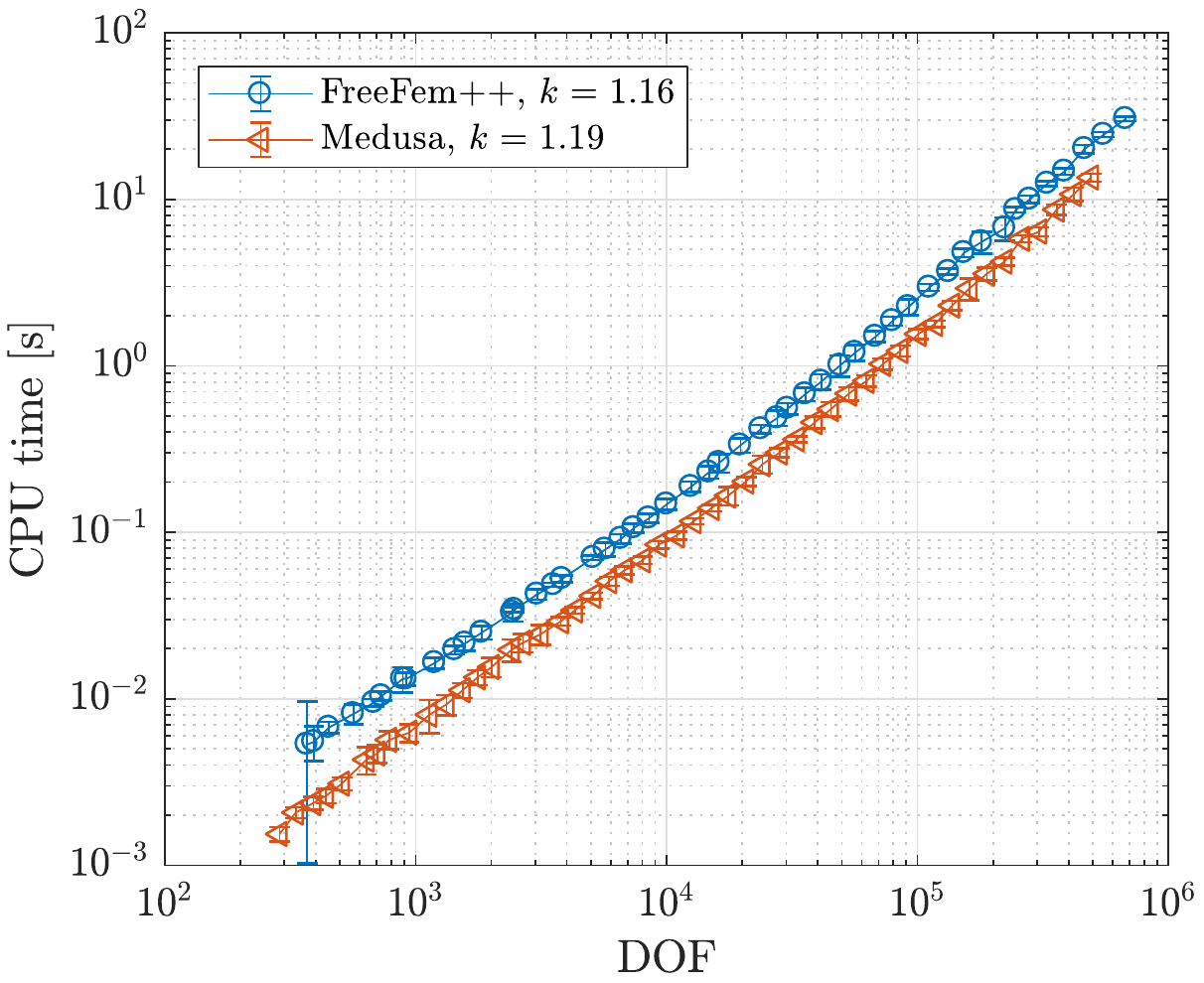}
  \includegraphics[width=0.49\linewidth]{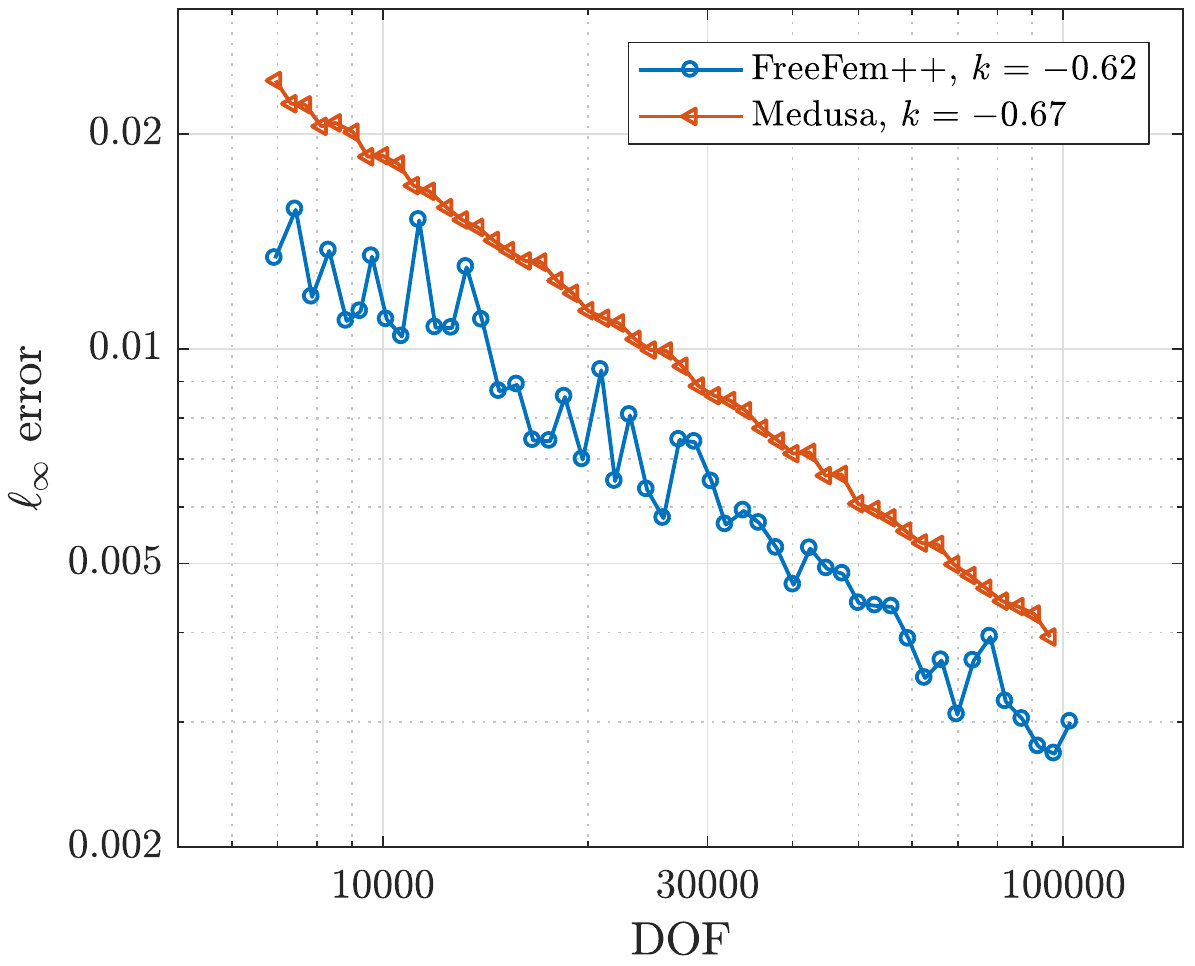}
  \includegraphics[width=0.49\linewidth]{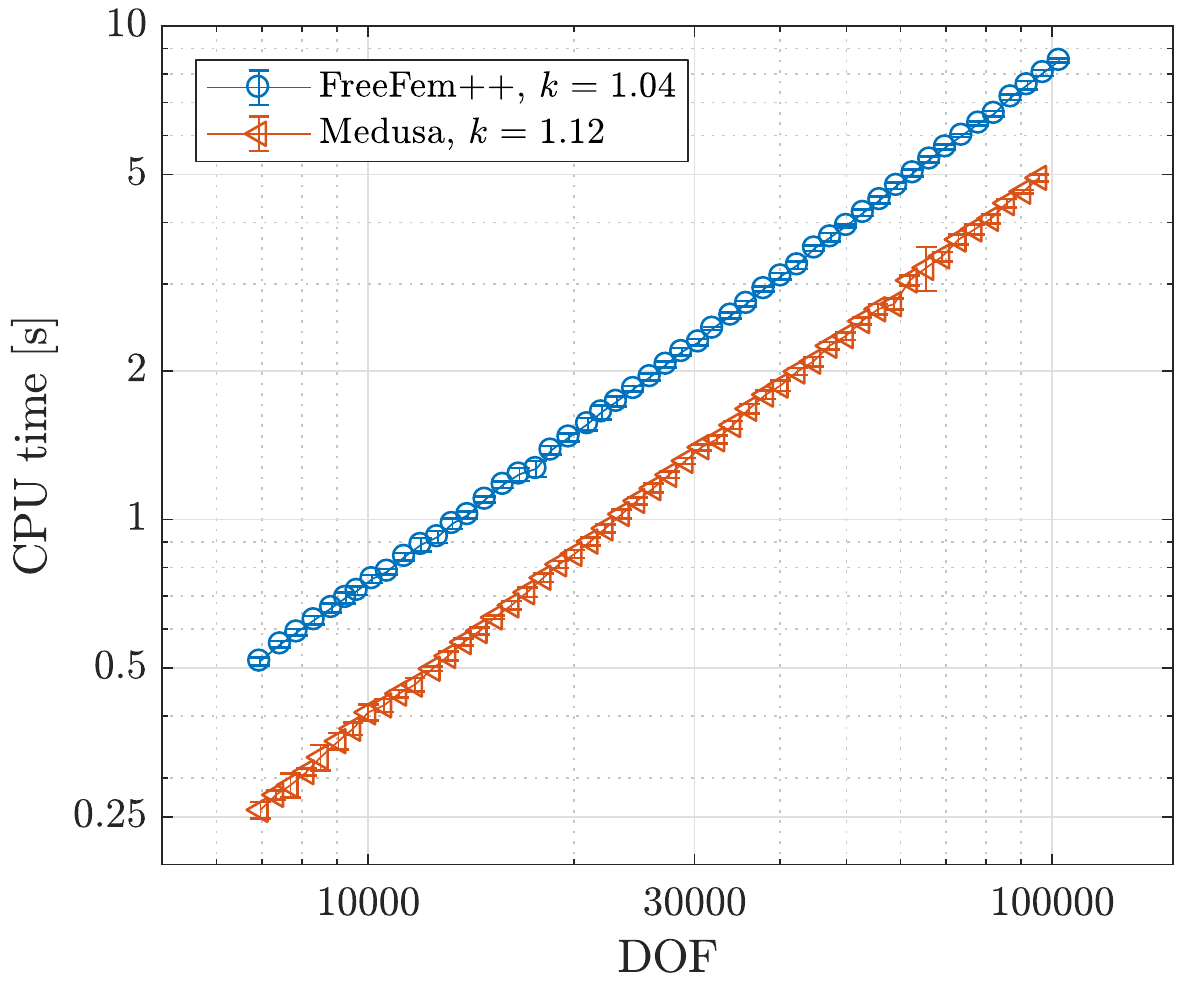}
  \caption{Errors and execution times of FreeFem++ and Medusa when solving~\eqref{eq:time-prob} in
  2D and 3D. Each time measurement was repeated 9 times. The median value is shown with error
  bars representing the standard deviation.}
  \label{fig:times}
\end{figure}

Both methods attain expected convergence rate $N^{-2/d}$ and similar accuracy, with
RBF-FD performing slightly worse.
The difference in execution times is almost exclusively due to node placing in Medusa
being faster than meshing in FreeFem++. The execution time is also highly dependent on the
number of stencil nodes, which can be lowered or increased, and on the choice of
sparse linear solver and its parameters. The Conjugate Gradient solver was chosen in FreeFem
because it performed best, and BiCGStab with ILUT$(5, 10^{-2})$  preconditioner was
chosen for Medusa. The solvers took approximately the same amount of time.

\begin{figure}
  \includegraphics[width=0.49\linewidth]{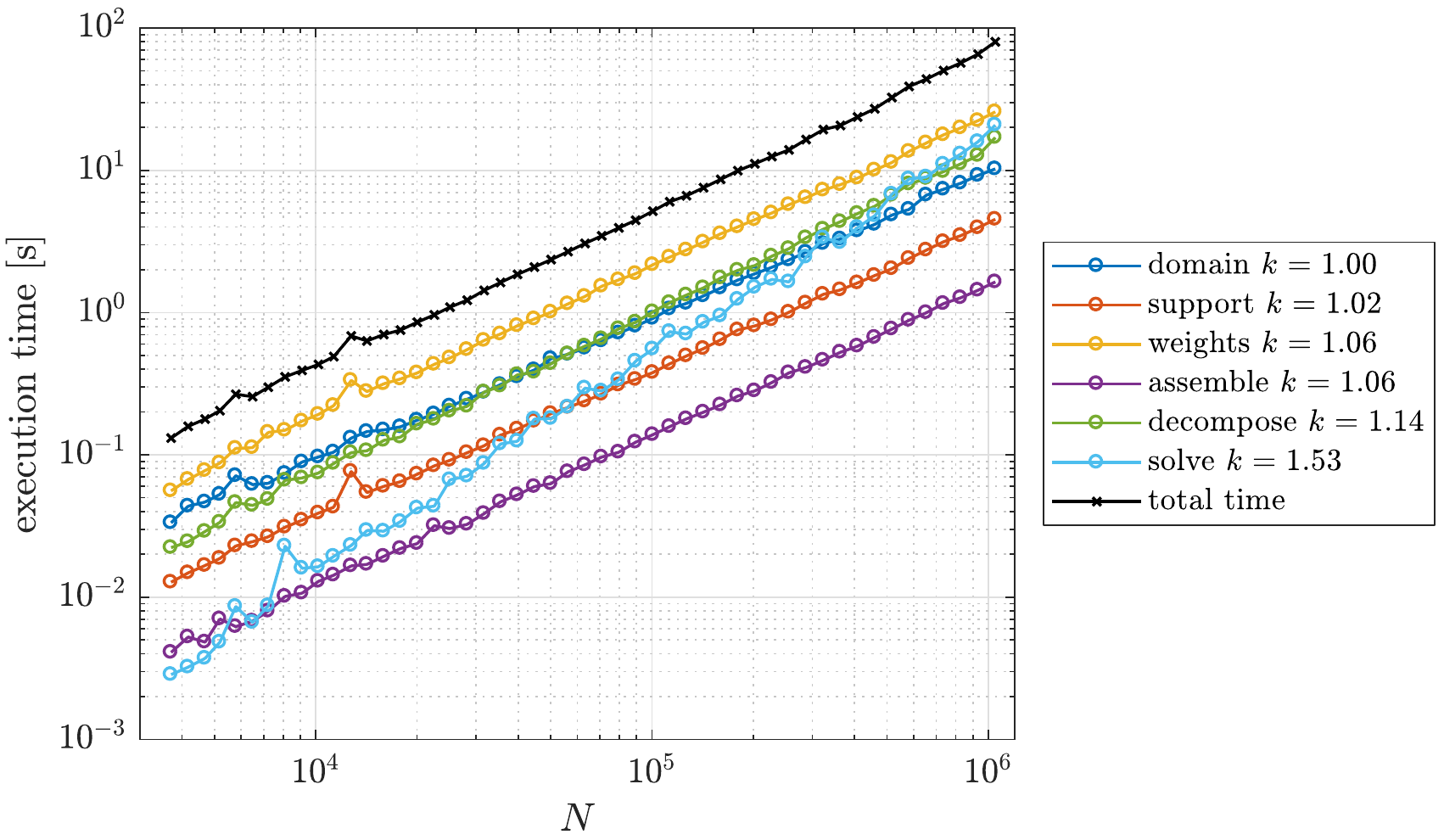}
  \includegraphics[width=0.49\linewidth]{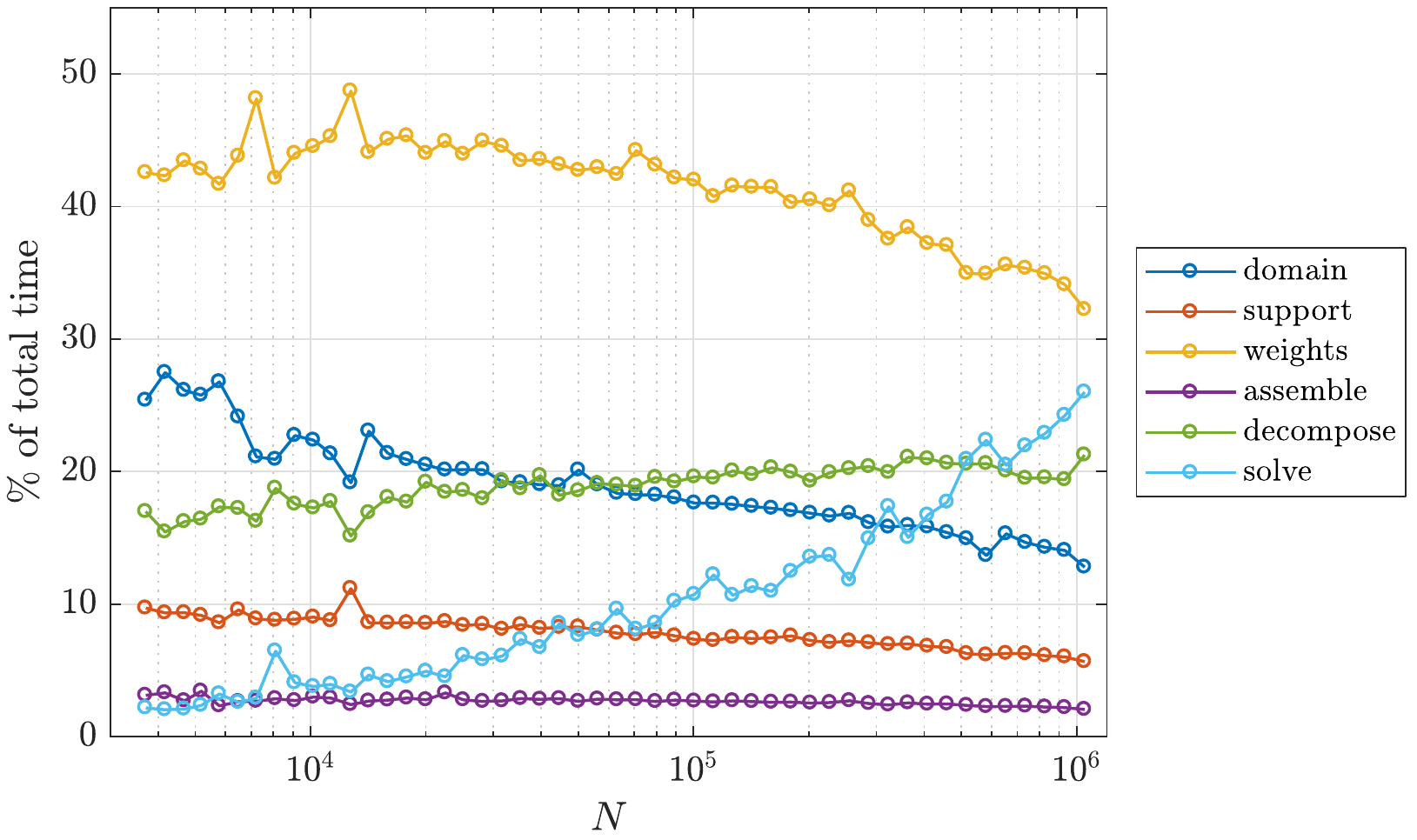}
  \caption{Errors and execution times of FreeFem++ and Medusa when solving~\eqref{eq:time-prob} in
    2D and 3D.}
  \label{fig:times-parts}
\end{figure}

Parts of the Medusa solution procedure were also timed separately: namely
domain discretization, stencil selection, stencil weight computation, matrix assembly,
preconditioner computation, iterative solution and post-process error computation.
Figure~\ref{fig:times-parts} shows these times with respect to the number of nodes
and a ratio of time spent on each part of the solution procedure.
These measurements also show the scaling behavior of different parts of the solution procedure.
Computational time complexity of most parts is linear or log linear, with the
exception of the linear solver. For most problems with explicit time iteration, the iteration
itself is so time consuming that domain discretization and weight computation are negligible, since
they are only performed once at the beginning of the iteration.

Execution time ratio can vary significantly in different setups. For 2D problems with 2nd order
methods, construction of domain discretization can take more than 50\% of the total time.
For high order methods with large support sizes and augmentation orders, weight computation
can severely dominate, even as high as 80\%. For more complicated problems and larger $N$,
linear solver can take up almost 90\% of the time.

These separate time measurements also serve as a guideline for optimization and parallelization.
Weight computation is trivially parallelizable and is already included in Medusa for shared memory
architectures, using OpenMP. Support for parallel sparse solvers is also included in Eigen,
and other parallelization efforts are ongoing.

Additionally, we also reviewed the cost of abstractions in performance critical sections by
comparing execution time with a ``bare-bones'' implementation~\cite{slak2018mipro} and
by analyzing assembly instructions with Compiler Explorer~\cite{compilerExplorer}, until
we were satisfied with incurred costs, which are now small to negligible.

\section{Conclusions and outlook}
\label{sec:conc}
In this paper we presented an overview of abstractions and implementation
of Medusa, a general purpose C++ library for solving PDEs with strong-form methods.
The library provides core elements of meshless solution procedures as standalone
blocks that can be pieced together or swapped to ease research, development and testing of
meshless methods, all in a dimension independent manner.
It allows to define custom node generation and stencil selection procedures,
basis functions, weights functions, RBFs, approximation schemes, and linear operators,
relying heavily on C++ templating system and most commonly used classes
are explicitly instantiated to avoid long compile times.
We have demonstrated this modularity and extensibility by constructing several
reported mesh-free methods and many more examples are available in the documentation.
Special attention is also paid to readability of the resulting code, which closely
resembles the mathematical description of the problem and allows the user to think in
terms of operators and fields instead of arrays and indices.
The library is also tested for correctness with a suite of unit tests and offers technical
documentation and other informal discussions on its website. A basic comparison of Medusa with
FreeFem++ on a Poisson problem showed it is comparable in execution time for similar accuracy.

Although Medusa is primarily intended as a research platform
for mesh-free community, it offers enough features for
solving 3D coupled problems, such as illustrated thermo-fluid transport problem
in an irregular 3D domain. Other problems, such as linear elasticity,
complex-valued electromagnetic scattering and wave propagation are also included in the examples.

The ongoing and future development of Medusa is aimed in several directions.
One is to increase the geometric capabilities of Medusa, by
adding a module for discretization of parametric surfaces,
and potentially extending it to handle Computer-Aided Design objects,
pushing Medusa a step closer to the engineering simulation software.

Another important directions is parallelism, since at the moment only
naive shared memory parallelization of modules that are trivial to execute in
parallel is offered. We are developing a parallel version of node positioning algorithms
as well as a domain decomposition module required for distributed parallel execution.

Throughout all other development we will also (albeit conservatively) extend the set of
approximations, bases, node generation algorithms and other elements offered by default,
with useful developments from ongoing research in core meshless areas. Potential future
additions include better support for adaptivity and coupled problems.

\begin{acks}
The authors would like to acknowledge other contributors to the Medusa library
(and its previous unpublished versions), listed in alphabetical order:
Urban Duh,
Mitja Jančič,
Maks Kolman,
Jure Lapajne,
Jure Močnik - Berljavac,
Anja Petković,
Anja Pirnat,
Ivan Pribec,
Tjaž Silovšek and
Blaz Stojanovič.

The authors would also like to acknowledge the financial support of the Slovenian Research
Agency (ARRS) research core funding No.\ P2-0095 and the Young Researcher program PR-08346.
\end{acks}

\bibliographystyle{ACM-Reference-Format}
\bibliography{references}

\end{document}